\documentclass[twocolumn,nofootinbib]{revtex4-1}  

\usepackage{dsfont}
\usepackage{rotating}
\usepackage{floatpag}
\rotfloatpagestyle{empty}

\usepackage{amsmath,amsthm,amssymb}
\usepackage{xspace,enumerate,epsfig}
\usepackage[dvipsnames]{xcolor}
\usepackage{graphicx}
\usepackage{marvosym}
\usepackage{stmaryrd}
\usepackage{docmute}
\usepackage{keycommand}
\usepackage{enumitem}

\usepackage[left=2.0cm,right=2.0cm,top=2.0cm,bottom=2.0cm]{geometry}

\newtheoremstyle{mytheoremstyle} 
    {\topsep}                    
    {\topsep}                    
    {}                   		  
    {}                           
    {\itshape}                   
    {:}                          
    {.5em}                       
    {}  
\theoremstyle{mytheoremstyle}

\newtheorem{theorem}{Theorem}[section]

\newtheorem{lemma}{Lemma}
\newtheorem{proposition}{Proposition}
\newtheorem{hypothesis}{Hypothesis}
\newtheorem{definition}{Definition}

\usepackage[utf8]{inputenc} 
\usepackage[english]{babel} 
\usepackage{verbatim}  
\usepackage{dsfont}
\usepackage[T1]{fontenc} 
\usepackage{amstext} 
\usepackage{ae}                

\usepackage{amsmath}
\allowdisplaybreaks		
\usepackage{mathtools,cancel}
\usepackage{amsfonts}
\usepackage{amssymb}
\usepackage{slashed}
\usepackage{xfrac} 
\usepackage{empheq}   
\usepackage{braket}
\usepackage{chngcntr}

\usepackage{graphicx}
\usepackage{epstopdf}
\usepackage[font=small]{caption}
\usepackage{subcaption}  
\usepackage[rightcaption]{sidecap}
\usepackage{wrapfig} 
\usepackage{lipsum} 
\usepackage{mdframed} 
\usepackage{tabularx}
\usepackage{url}
\usepackage{hhline} 	
\usepackage{fancyhdr}   
\usepackage{varwidth} 
\usepackage[absolute,overlay]{textpos}
\usepackage{array}            
\usepackage{float}	
\usepackage{booktabs}
\usepackage{dashbox}  
\usepackage{breqn}    
\usepackage{hyperref}

\counterwithout{theorem}{section}

\newcommand{\Trace}{\textnormal{Tr}}

\begin{document}

\title{Cyclic Quantum Causal Models}
\author{Jonathan Barrett$^1$, Robin Lorenz$^{1,2}$, Ognyan Oreshkov$^3$}
\affiliation{$ ^1$Department of Computer Science$,$ University of Oxford$,$ Wolfson Building$,$ Parks Road$,$ Oxford OX1 3QD$,$ UK\\
$ ^2$Cambridge Quantum Computing Ltd$,$ 9a Bridge Street$,$ Cambridge$,$ UK\\
$ ^3$QuIC$,$ Ecole Polytechnique de Bruxelles$,$ C.P. 165$,$ Universit\'{e} Libre de Bruxelles$,$ 1050 Brussels$,$ Belgium}

\begin{abstract}

Causal reasoning is essential to science, yet quantum theory challenges it. Quantum correlations violating Bell inequalities defy satisfactory causal explanations within the framework of classical causal models. What is more, a theory encompassing quantum systems and gravity is expected to allow causally nonseparable processes featuring operations in indefinite causal order, defying that events be causally ordered at all. The first challenge has been addressed through the recent development of intrinsically quantum causal models, allowing causal explanations of quantum processes -- provided they admit a definite causal order, i.e. have an acyclic causal structure. 
This work addresses causally nonseparable processes and offers a causal perspective on them through extending quantum causal models to cyclic causal structures. Among other applications of the approach, it is shown that all unitarily extendible bipartite processes are causally separable and that for unitary processes, causal nonseparability and cyclicity of their causal structure are equivalent.
\end{abstract}

\maketitle

\section*{INTRODUCTION}

\let\thefootnote\relax\footnotetext{This is a post-peer-review, pre-copyedit version of an article published in \textit{Nature Communications} 12, 885 (2021). The final authenticated version is available online
at: \url{https://doi.org/10.1038/s41467-020-20456-x}.}

There has been growing interest in higher-order quantum processes in which separate operations do not occur in a definite causal order (see, e.g., Refs.~\cite{Hardy_2005_ProbabilityTheoriesWithDynamicCausalStructure, 
ChiribellaEtAl_2013_QuantumCompWithoutDefCausalStructure, 
OreshkovEtAl_2012_QuantumCorrelationsWithoutCausalOrder, 
Chiribella_2012_PerfectDiscriminationOfChannelsViaSuperpositionCS, 
AraujoEtAl_2014_ComputationalAdvantage, 
AraujoEtAl_2015_WitnessingCausalNonSeparability, 
OreshkovEtAl_2016_CausallySeparableProcesses, 
GuerinEtAl_2016_ExponentialCommunicationComplexityAdvantage, 
BranciardEtAl_2015_SimplestCausalInequalitiesAndViolations, 
OreshkovEtAl_2016_OperationalQuantumTheoryWithoutTime, 
Baumeler_2017_ReversibleTimeTravelWithFreedomOfChoice, 
Baumeler_EtAl_2016_SpaceOfLOgicallyConsistentClassicalProcesses, 
SilvaEtAl_2017_ConnectingIndefiniteWithMultiTime, 
AbbottEtAl_2017_GenuinelyMultipartiteNoncausality, 
Portmann_2017_CausalBoxes, 
MiklinEtAl_2017_EntropicApproachToCausalCorrelations, 
JiaEtAl_2018_TensorProductsOfProcessMatricesWithIndefiniteCS, 
Oreshkov_2019_TimeDelocalizedSubsystems, 
EblerEtAl_2018_EnhancedCommunicationFromIndefiniteCausalOrder,  
CastroEtAl_2018_DynamicsOfQuantumCausalStructures, 
UijlenEtAl_2019_CategoricalSemanticsCausalStructure, 
TobarEtAl_2020_ReversibleDynamicsWithCTCs} for a selection). 
This property, called causal nonseparability  
\cite{OreshkovEtAl_2012_QuantumCorrelationsWithoutCausalOrder, 
AraujoEtAl_2015_WitnessingCausalNonSeparability, 
OreshkovEtAl_2016_CausallySeparableProcesses, WechsEtAl_2019_MultipartiteCausalNonSeparability} 
was formalized within the process matrix framework \cite{OreshkovEtAl_2012_QuantumCorrelationsWithoutCausalOrder}, 
which describes correlations between quantum nodes of intervention without assuming a predefined order between the nodes. Challenging conventional notions of causality, causally nonseparable processes have been shown to allow informational tasks that cannot be achieved with operations used in a definite order \cite{Chiribella_2012_PerfectDiscriminationOfChannelsViaSuperpositionCS,
AraujoEtAl_2014_ComputationalAdvantage,  
FeixEtAl_2015_QuantumSuperpositionOfOrderAsResource, 
GuerinEtAl_2016_ExponentialCommunicationComplexityAdvantage}. 
Such processes have been conjectured to be relevant in the context of quantum gravity \cite{Hardy_2005_ProbabilityTheoriesWithDynamicCausalStructure, ChiribellaEtAl_2013_QuantumCompWithoutDefCausalStructure, OreshkovEtAl_2012_QuantumCorrelationsWithoutCausalOrder, ZychEtAl_2019_BellsTheoremForTemporalOrder} 
and closed time-like curves \cite{ChiribellaEtAl_2013_QuantumCompWithoutDefCausalStructure, OreshkovEtAl_2012_QuantumCorrelationsWithoutCausalOrder, Brukner_QuantumCausality, AraujoEtAl_2017_QuantumComputationWithIndefiniteCausalStructure, Baumeler_2017_ReversibleTimeTravelWithFreedomOfChoice, TobarEtAl_2020_ReversibleDynamicsWithCTCs}, but some are also known to admit realizations in standard quantum mechanics on time-delocalized systems \cite{Oreshkov_2019_TimeDelocalizedSubsystems}. 
A prominent example is the quantum SWITCH, which has been demonstrated experimentally \cite{ProcopioEtAl_2015_ExperimentalSuperpositionOfOrders, RubinoEtAl_2017_ExperimentalVerificationIndefiniteCausalOrders, GoswamiEtAl_2018_IndefiniteCausalOrderInQuantumSWITCH, RubinoEtAl_2019_ExperimentalEntanglementOfTemporalOrders, WeiEtAl_2019_ExperimentalQuantumSWITCHCommunicationComplexity, GuoEtAl_2020_ExperimentalTransmitionOfQIThroughSuperpositionOfCausalOrder}. 

On a separate front, there is the recent development of the framework of quantum causal models \cite{AllenEtAl_2016_QCM, BarrettEtAl_2019_QCMs} (see, e.g., Refs.~\cite{Tucci_1995_QuantumBayesianNets,  
Leifer_2006_QuantumDynamicsAndConditionalProbability, 
LeiferEtAl_2008QuantumGraphicalModels,
Laskey_2007_QuantumCausalNetworks, 
LeiferEtAl_2013_QTAsBayesianInference, 
HensonEtAl_2014_TheoryIndependentLimitsGeneralisedBayesianNetworks, 
PienaarEtAl_2015_GraphSeparationTheoremForQCMs, 
RiedEtAl_2015_QuantumAdvantageForCausalInference,
Fritz_2016_BeyondBellsTheorem,
CostaEtAl_2016_QuantumCausalModeling,
Pienaar_2018_QuantumCausalModelsViaQuantumBayesianism} for related, previous work) as a fully quantum version of the classical framework of causal models \cite{Pearl_Causality, SpirtesEtAL_2000_BookCausationPredictionSearch}. 
It is formulated within the formalism of process matrices, but contains the classical causal models as special cases and generalizes many of the fundamental concepts and core theorems of the latter. 
Quantum causal models thus constitute a general framework for reasoning about quantum systems in causal terms, allowing the rigorous study of the empirical constraints imposed by quantum causal structures 
-- however, only as far as causal structures are concerned that are expressible as directed acyclic graphs (DAGs), i.e., where there is a well-defined causal order. 
The central idea behind the approach in Refs.~\cite{AllenEtAl_2016_QCM, BarrettEtAl_2019_QCMs} is that causal relations between quantum systems, as encoded in a DAG, correspond to influence through underlying unitary transformations. This facilitated, in particular, a justification of the quantum Markov condition relative to a DAG that underpins the definition of a quantum causal model -- any such model can be thought of as arising from a unitary circuit fragment with a compatible causal structure by marginalizing over latent local disturbances \cite{BarrettEtAl_2019_QCMs}. 

It is a natural question whether these hitherto separate lines of research can be merged to arrive at a causal model perspective on processes that are not compatible with a fixed order of the quantum nodes. 
While this direction of thought has been considered in earlier work (see, e.g., Refs.~\cite{CostaEtAl_2016_QuantumCausalModeling, MacleanEtAl_2017_QuantumCoherentMixturesOfCausalRelations, GiarmatziEtAl_2018_CausalDiscoveryAlgorithm}), it was previously not clear how to take the idea forward due to various conceptual and technical obstacles -- including, for example, how quantum nodes and the quantum Markov condition should be defined, how the notion of autonomy of causal mechanisms should be understood  \cite{MacleanEtAl_2017_QuantumCoherentMixturesOfCausalRelations}, and how to prevent paradoxes. 

This work overcomes these obstacles by generalizing the approach to quantum causal models of Refs.~\cite{AllenEtAl_2016_QCM, BarrettEtAl_2019_QCMs}. 
A large class of processes that are not compatible with a fixed order of the nodes can then be understood to have a causal structure, albeit one that includes directed cycles. 
This may appear counterintuitive, but the process matrix framework guarantees that it is free of paradoxes.  
The motivation for entertaining such a proposal is two-fold. First, in light of the puzzling nature of causally nonseparable processes and the open question of which ones are physically possible in nature, a conceptual clarification of causal structure is an important next step. Second, our approach yields mathematical tools facilitating new technical results, including a more fine-grained description of the compositional structure of a process that is implied by its causal properties.
One of the implications of the latter derived in this work is a proof that all bipartite processes that admit a unitary extension \cite{AraujoEtAl_2017_PurificationPostulateForQMWithIndefiniteCausalOrder} are causally separable. We also prove that for unitary processes, causal nonseparability and cyclicity of their causal structure are equivalent.

\section*{RESULTS}

\subsection*{The process formalism, causal order and signalling} 

Let us start by setting out some necessary background and essential concepts. 
In quantum theory, a system $A$ is associated with a complex Hilbert space $\mathcal{H}_A$, and its state is a density operator $\rho_A \in \mathcal{L}(\mathcal{H}_A)$, where $\mathcal{L}(\mathcal{H}_A)$ is the space of linear operators over $\mathcal{H}_A$. The most general evolution of a system, assuming that it is initially uncorrelated with its environment, is given by a completely positive trace preserving (CPTP) map $\mathcal{E}: \mathcal{L}(\mathcal{H}_A) \rightarrow \mathcal{L}(\mathcal{H}_B)$, where this notation allows the output system to be different from the input system.  
The most general operation that an agent can perform from an input system A to an output system B has a classical outcome $k$ and specifies the transformation from A to B conditioned on each value of $k$ being obtained. Mathematically, the operation corresponds to a quantum instrument, which is a collection of completely positive (CP) maps $\{\mathcal{E}^{k}:\mathcal{L}(\mathcal{H}_{A})\rightarrow \mathcal{L}(\mathcal{H}_{B})\}$, such that $\mathcal{E} = \sum_{k} \mathcal{E}^{k}$ is a trace-preserving CP map. 

It is convenient to represent CP maps with operators, via a variant of the Choi-Jamio\l kowski (CJ) isomorphism \cite{Jamolkowski_1972, Choi_1975}, which to a given CP map  $\mathcal{E}: \mathcal{L}(\mathcal{H}_A) \rightarrow \mathcal{L}(\mathcal{H}_B)$ associates the CJ operator $\rho^{\mathcal{E}}_{B|A} := \sum_{i,j} \ \mathcal{E}(\ket{i}_A\bra{j}) \otimes  \ket{i}_{A^*}\bra{j}$, where $\left\{ \ket{i}_A \right\}$ is an orthonormal basis of $\mathcal{H}_A$, and $\left\{ \ket{i}_{A^*} \right\}$ the corresponding dual basis. The CJ operator for a CPTP map  $\mathcal{E}$ satisfies $\Trace_{B}[\rho^{\mathcal{E}}_{B|A}] = \mathds{1}_{A^*}$. This variant of the CJ isomorphism is used in Refs.~\cite{AllenEtAl_2016_QCM, BarrettEtAl_2019_QCMs}, and has the advantage that the CJ operator is both positive semi-definite and independent of the basis used in its definition. 

The idea behind the process formalism \cite{OreshkovEtAl_2012_QuantumCorrelationsWithoutCausalOrder} is that there is a fixed set of locations $A_i$, $i=1,\cdots, n$, which in this work we call quantum nodes, at each of which an agent can perform an operation on a quantum system. A quantum node $A_i$ is associated with two Hilbert spaces, an input Hilbert space $\mathcal{H}_{A_i^{\textrm{in}}}$ and an output Hilbert space $\mathcal{H}_{A_i^{\textrm{out}}}$ (both here assumed finite-dimensional). The input Hilbert space carries the state of the input system just before the operation by the agent, and the output Hilbert space carries the state of the output system just after the operation. The operation itself corresponds to a quantum instrument $\{\mathcal{E}_A^{k_A}:\mathcal{L}(\mathcal{H}_{A^{\textrm{in}}})\rightarrow \mathcal{L}(\mathcal{H}_{A_{\textrm{out}}})\}$. Conceptually, a quantum node is sometimes  thought of as representing a small, localized laboratory in some region of spacetime, but may also be conceived more abstractly, for example as occupying a particular position in between the gates of a quantum circuit.

The aim of the process formalism is to describe the correlations between the outcomes of the operations that are performed at the separate quantum nodes. Given a set of instruments at the quantum nodes $A_1,...,A_n$, the joint probability for their outcomes is given by 
\begin{equation}\label{Eq_quantumprocesstracerule}
P(k_{A_1}, \ldots , k_{A_n}) = \Trace \Big[ \sigma_{A_1 ... A_n} \Big( \bigotimes_i \tau^{k_{A_i}}_{A_i} \Big)  \Big] \ , 
\end{equation} 
where $\tau^{k_A}_A := \left( \rho^{\mathcal{E}^{k_A}}_{{A^{\text{out}}}|A^{\text{in}}} \right)^T$, and 
$\sigma_{A_1 ... A_n} \in \mathcal{L}( \bigotimes_i \mathcal{H}_{A^{\text{in}}_i} \otimes \mathcal{H}^*_{A^{\text{out}}_i})$ is called the process operator, and which we will also sometimes refer to more simply as the process. 

A process operator $\sigma_{A_1 ... A_n}$, which up to a different convention of the CJ isomorphism is the same as a process matrix \cite{OreshkovEtAl_2012_QuantumCorrelationsWithoutCausalOrder}, obeys constraints, designed to ensure that valid joint probabilities are returned by Eq.~(\ref{Eq_quantumprocesstracerule}) for any possible choices of the operations performed by the agents, and that the same holds even when the agents have pre-shared entanglement. These constraints are  \cite{OreshkovEtAl_2012_QuantumCorrelationsWithoutCausalOrder}: $\sigma_{A_1 ... A_n} \geq 0$ and $\Trace [ \sigma_{A_1 ... A_n} \ (\tau_{A_1} \otimes \cdots \otimes \tau_{A_n}) ] = 1$, for any set of CPTP maps $\{\tau_{A_i}\}$ at the $n$ nodes. Simple-to-check necessary and sufficient conditions for an operator in $\mathcal{L}( \bigotimes_i \mathcal{H}_{A^{\text{in}}_i} \otimes \mathcal{H}^*_{A^{\text{out}}_i})$ to be a valid process operator can be found in Refs.~\cite{AraujoEtAl_2015_WitnessingCausalNonSeparability, OreshkovEtAl_2016_CausallySeparableProcesses}. To avoid clutter when tracing over a node $A$ we will write $\Trace_{A}[ \rule{0.25cm}{0.4pt}] := \Trace_{A^{\text{in}}(A^{\text{out}})^*}[ \rule{0.25cm}{0.4pt}]$.

A question of central interest in the study of the process formalism has been whether a given process operator is compatible with the existence of a definite causal order of its nodes. A closely related question concerns how this relates to the possibilities for signalling between different nodes. Let us first make these notions more precise.

\begin{figure}[H]
	\centering
	\vspace*{-0.2cm}
	\includegraphics[scale=1.0]{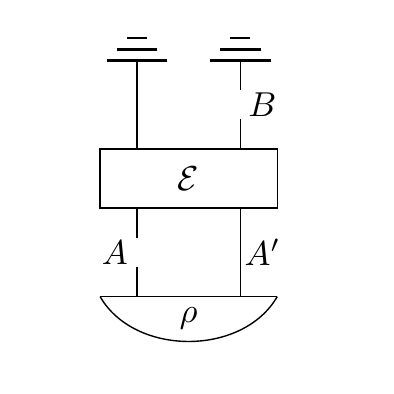}
	\vspace*{-0.3cm}
	\caption{Simple circuit with two quantum nodes $A$ and $B$. \label{Fig_Definitecausalorder}}
\end{figure}

Consider the sequence of quantum operations represented in the form of a circuit in Fig.~\ref{Fig_Definitecausalorder}. 
The gate in the circuit corresponds to an arbitrary CPTP map $\mathcal{E}$, and the initial preparation to an arbitrary bipartite state $\rho$. Quantum nodes $A$ and $B$ correspond to positions in the circuit in between gates, at which an agent can choose to perform a quantum instrument on the system at that position. The nodes are represented as broken wires, with it understood that the agent's instrument mediates the two pieces. The lower piece of the wire corresponds to the input Hilbert space of the quantum node and the upper piece of the wire to the output Hilbert space. Any circuit with some wires broken defines a partial order over the quantum nodes, with a node $N$ preceding node $N'$ in the partial order if and only if there is a path from $N$ to $N'$ along the (broken or unbroken) wires of the circuit. We call this partial order the causal order. In the example, node $A$ precedes node $B$ in the causal order. The circuit defines joint probabilities for the outcomes of any quantum instruments that are performed at nodes $A$ and $B$, hence defines a process operator over the nodes $A$ and $B$. 

An important concept is now that of causal separability, first introduced in Ref.~\cite{OreshkovEtAl_2012_QuantumCorrelationsWithoutCausalOrder} for the bipartite case. 
A bipartite process $\sigma_{AB}$ is called causally separable iff it can be seen to arise as a convex mixture of processes with a fixed causal order between $A$ and $B$, i.e., $\sigma_{AB} = p \ \sigma_{AB}^{A \npreceq B} + (1-p) \ \sigma_{AB}^{B \npreceq  A}$, 
with $0 \leq p \leq 1$, where $\sigma_{AB}^{B \npreceq  A}$ is  a process that can arise from a sequence of operations of the form of Fig.~\ref{Fig_Definitecausalorder}, and $\sigma_{AB}^{A \npreceq  B}$ is a process that can arise from a different sequence of operations, of the same form except that $B$ precedes $A$. Otherwise $\sigma_{AB}$ is causally nonseparable. The idea is that a causally separable process can be thought to describe a situation in which a well defined, though possibly unknown, causal order of the nodes exists, whereas a causally nonseparable process is not compatible with such an interpretation.

The connection with signalling between the nodes is as follows. Given a bipartite process $\sigma_{AB}$, we say that there is no signalling from quantum node $B$ to quantum node $A$ if and only if for all quantum instruments $\tau_A^{k_A}$ at $A$ and all deterministic quantum instruments $\tau_B$ at $B$, the probability distribution $P(k_A)=\Trace [\sigma_{AB} (\tau_A^{k_A} \otimes \tau_B )]$ is independent of $\tau_B$. This condition is equivalent to $\sigma_{AB} = \sigma_{A B^{\text{in}}  } \otimes \mathds{1}_{(B^{\text{out}})^*}$  \cite{OreshkovEtAl_2012_QuantumCorrelationsWithoutCausalOrder}. The connection between signalling and causal order is that in any sequence of operations of the form of Fig.~\ref{Fig_Definitecausalorder}, there is no signalling from $B$ to $A$. Moreover, every bipartite process operator with no-signalling from $B$ to $A$ is known to have a realisation as the process operator arising from a circuit of the form of Fig.~\ref{Fig_Definitecausalorder} \cite{ChirbiellaEtAl_2009_QuantumNetworkFramework}.

The formal definition of the multipartite generalization of causal separability is more intricate than in the bipartite case: beyond just convex mixtures of fixed causal orders, the definition allows for a dynamical causal order in which the causal order at later quantum nodes can depend on the events taking place at earlier quantum nodes. This definition is postponed to a later section dedicated to causal separability. See also Refs.~\cite{OreshkovEtAl_2016_CausallySeparableProcesses, WechsEtAl_2019_MultipartiteCausalNonSeparability} for a detailed discussion. A discussion of signalling in a multipartite process is also more involved, since whether a subset of quantum nodes can signal to another subset of quantum nodes depends on the interventions performed at other quantum nodes not in the two subsets.

\subsection*{Causal influence vs signalling} 

This work is concerned with a notion of causal structure, which is distinct from the causal order defined by a circuit, and which also needs to be carefully distinguished from the possibilities for signalling afforded by a general process operator. In order to motivate the idea, 
consider a circuit of the form of Fig.~\ref{Fig_Definitecausalorder}, with each wire representing a qubit, with $\rho = \rho_{A^{\text{in}}} \otimes |0\rangle_{A'} \langle 0|$, and with the channel $\mathcal{E}$  being a quantum Controlled-NOT gate with the control on the output wire of the $A$ node. This circuit defines a process operator $\sigma^0_{AB}$ on the $A$ and $B$ nodes, which may easily be computed, and it can be verified that 
$\sigma^0_{AB}$ allows signalling from $A$ to $B$. Similarly, in the same circuit except with $\rho = \rho_{A^{\text{in}}} \otimes |1\rangle_{A'} \langle 1|$, the process operator $\sigma^1_{AB}$ is easily computed, and it can be verified that signalling is possible from $A$ to $B$. 

Now consider the same experiment, except with the preparation of the $A'$ system given by flipping a fair coin, and preparing $|0\rangle_{A'} \langle 0|$ on heads and $|1\rangle_{A'} \langle 1|$ on tails. If the outcome of the coin flip is unknown, then the state of the $A'$ system is the mixed state $\mathds{1}/2$, and the corresponding process operator over $A$ and $B$ is 
\begin{equation}
\sigma^{\mathrm{mix}}_{AB} = \mathds{1}_{(B^{\text{out}})^*} \otimes (1/2) \mathds{1}_{B^{\text{in}}} \otimes \mathds{1}_{(A^{\text{out}})^*} \otimes \rho_{A^{\text{in}}}.
\end{equation}
In the process $\sigma^{\mathrm{mix}}_{AB}$, there is no signalling from $A$ to $B$. Indeed the very same process operator could arise from a situation in which $A$ and $B$ are independent and spacelike separated.

In the experiment with the coin flip, it is clear that $A$ has a causal influence on $B$, since agents who know the value of the coin flip would be able to send signals from $A$ to $B$. From the perspective of agents who do not know the value of the coin flip, however, signalling is washed out by the randomness of the unobserved system. 
A similar phenomenon is well understood in the literature on classical causal modelling. In a canonical example, $A$, $B$, and $C$ are all classical bits, with $A$ and $C$ causes of $B$ such that $B$ is equal to the parity of $A$ and $C$. If $C$ is inaccessible, or hidden, and satisfies $P(C=0)=P(C=1)=1/2$, then $B$ is randomly distributed regardless of the value of $A$. Hence as long as $C$ remains hidden, signals cannot be sent from $A$ to $B$. (See Section~2.4 of Ref.~\cite{BarrettEtAl_2019_QCMs}.)

The conclusion that should be drawn from the example with process $\sigma^{\mathrm{mix}}_{AB}$ is that causal influence between quantum nodes should not be defined in terms of the possibilities for signalling afforded by a process operator, at least not if there is a chance that unobserved systems ($A'$ in the example) are interacting with the systems under study ($A$ and $B$ in the example). Given only a process operator $\sigma_{AB}$ and no other data, although signalling is sufficient for causal influence, it can happen that $A$ has a causal influence upon $B$ even though there is no signalling from $A$ to $B$ in $\sigma_{AB}$.

\subsection*{Quantum causal models}

The framework of quantum causal models, introduced in Refs.~\cite{AllenEtAl_2016_QCM, BarrettEtAl_2019_QCMs}, is based on the idea that in an example like that just above, statements about causal influence can be defined in terms of signalling, but only once all relevant systems are included in the description. At this point, the description is of a closed system, and at least in standard quantum theory, evolution of a closed system is unitary. Hence quantum causal models define causal influence in terms of unitary transformations. Ref.~\cite{BarrettEtAl_2019_QCMs} shows that in the case of unitary circuits with broken wires representing quantum nodes, the causal relations between the quantum nodes can be summarized in the form of a DAG, where the DAG imposes constraints on the process operator over the quantum nodes.

This leaves open the question of what the pattern of causal influence might be in causally nonseparable processes described in the literature. Can it even be well defined or must one conclude that these processes are not amenable to causal explanation at all, or that all that can be discussed is signalling between the nodes? Our idea is that such processes can be understood in causal terms, if the framework of quantum causal modelling is extended to allow causal cycles. We will show that the resulting formalism can be used successfully to describe some of the much-studied instances of causally nonseparable processes from the literature. Later, we show the utility of this approach by using it to settle previously open questions concerning causally nonseparable processes.

The following definition generalizes that of Refs.~\cite{AllenEtAl_2016_QCM, BarrettEtAl_2019_QCMs}, by allowing cyclic graphs (along with a more minor generalization, which is that the input and output Hilbert spaces of a quantum node can here have different dimensions).

\begin{definition} \textnormal{(Quantum causal model (QCM) --- generalized)} \label{Def_QCM}
	A \textnormal{QCM} is given by:
\begin{enumerate}[label=\textnormal{(\arabic*)}, leftmargin=0.7cm]
\item a causal structure represented by a directed graph $G$ with vertices corresponding to quantum nodes $A_1, ... , A_n$, 
\item for each $A_i$, a quantum channel $\rho_{A_i | Pa(A_i)} \in \mathcal{L}(\mathcal{H}_{A_i^{\textrm{in}}} \otimes \mathcal{H}^*_{Pa(A_i)^{\textrm{out}}})$, where 
$Pa(A_i)$ denotes the set of parents of $A_i$ according to $G$, such that $[\rho_{A_i | Pa(A_i)} \ , \ \rho_{A_j | Pa(A_j)} ] \allowbreak = 0$ for all $i,j$ and such that $
\sigma_{A_1 ... A_n} = \prod_i \rho_{A_i | Pa(A_i)}$ is a process operator over the quantum nodes $A_1 , ... , A_n$.
\end{enumerate}
\end{definition}

When writing products of the form $\prod_i \rho_{A_i | Pa(A_i)}$, it is understood implicitly that each factor is padded  with an identity operator in tensor product for all other spaces.  
A QCM is called cyclic iff its causal structure contains directed cycles, and acyclic otherwise. 

It is useful to define a term to express the fact that a given process operator $\sigma$ has the correct form with respect to a given causal structure to define a QCM.   
\begin{definition} \textnormal{(Quantum Markov condition --- generalized)} \label{Def_MarkovCondition}
A process $\sigma_{A_1 ... A_n}$ is called \textnormal{Markov} for a directed graph $G$ with quantum nodes $A_1,\ldots, A_n$ as its vertices iff it admits a factorization into pairwise commuting channels of the form $\sigma_{A_1 ... A_n} = \prod_{i=1}^n \rho_{A_i | Pa(A_i) }$. 
\end{definition}

Note that the Markov condition of classical causal models \cite{Pearl_Causality, SpirtesEtAL_2000_BookCausationPredictionSearch} is a special case of Def.~\ref{Def_MarkovCondition}, obtained when the graph is acyclic and $\sigma_{A_1 ... A_n}$ is diagonal in a product basis, and encodes a classical probability distribution \cite{BarrettEtAl_2019_QCMs}. 
The following first sets out some further terminology and basic properties of Def.~\ref{Def_QCM} and then turns to motivating and explaining Def.~\ref{Def_QCM}, making the link with unitary transformations, and showing why it is that for a particular directed graph, condition (2) should hold. 

\begin{figure}[H]
	\centering
	\includegraphics[scale=1.0]{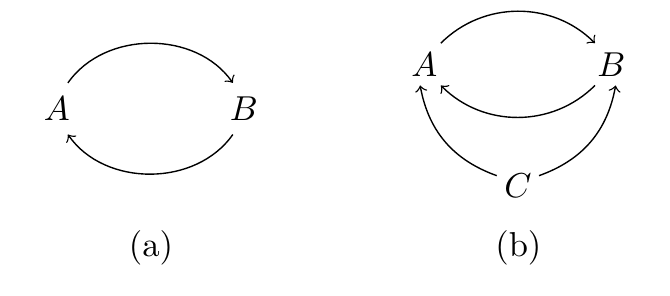}
	\vspace*{-0.3cm}
	\caption{Examples of cyclic directed graphs. \label{Fig_ExampleGraphs}}
\end{figure}

First, observe that not every cyclic graph supports a QCM in an interesting way. Consider, for example, the two-node cyclic graph of Fig.~2a. A QCM with such a causal structure would come with a process operator 
\begin{equation}
\sigma_{AB} = \rho_{A|B} \ \rho_{B|A} \ . \label{Eq_BipartiteCyclic}
\end{equation}
Here and throughout, channels between the nodes on which a process is defined are written such that anything appearing to the right of the bar refers to the output Hilbert space of the node, and anything appearing to the left of the bar refers to the input Hilbert space of the node. 
By our conventions $ \rho_{A|B} \ \rho_{B|A} = \rho_{A|B} \otimes \rho_{B|A}$. However, this is not a valid process operator unless either $\rho_{A|B} = \rho_{A^{\textrm{in}}} \otimes \mathds{1}_{(B^{\textrm{out}})^*}$, or $\rho_{B|A} =  \rho_{B^{\textrm{in}}} \otimes \mathds{1}_{(A^{\textrm{out}})^*}$. 
In other words, at least one of the channels $\rho_{A|B}$, $\rho_{B|A} $ carries no information, but simply ignores its input and prepares a fixed state on the output. Intuitively speaking, this is because there would otherwise be logical paradoxes for certain choices of interventions at $A$ and $B$.

More generally, we will say that a QCM is faithful iff each of the channels $\rho_{A_i | Pa(A_i)}$ is signalling from $A_j^{\textrm{out}}$ to $A_i^{\textrm{in}}$ for every $A_j\in Pa(A_i)$, i.e., 
\begin{equation}
\rho_{A_i | Pa(A_i)} \ne \frac{1}{d_j} \mathrm{Tr}_{(A_j^{\textrm{out}})^*} (\rho_{A_i | Pa(A_i)}) \otimes \mathds{1}_{(A_j^{\textrm{out}})^*},
\end{equation}
where $d_j$ is the dimension of $A_j^{\textrm{out}}$. 
Our claim concerning the causal structure of Fig.~2a can be summarized as:
\begin{proposition} \label{Prop_NoBipartite}
There is no faithful cyclic quantum causal model with two nodes.
\end{proposition}
\noindent \textit{Proof:} See Methods.

Now consider the cyclic graph $G'$ in Fig.~2b 
A QCM with $G'$ as its causal structure comes with the data 
\begin{equation}
\sigma_{ABC} = \rho_{A|BC} \ \rho_{B|AC} \ \rho_{C} \ . \label{Eq_TriartiteGenericQCM}	
\end{equation}
Eq.~\ref{Eq_TriartiteGenericQCM}, compared to Eq.~\ref{Eq_BipartiteCyclic}, has the key difference that the commuting operators have non-trivial action on $(C^{\textrm{out}})^*$. As a result, it turns out that 
faithful cyclic QCMs of this form do exist. An example is described below.

Note that, given a cyclic graph such as that in Fig.~2b, even when a faithful QCM exists it is not in general the case that any set of commuting channels $\rho_{A_i | Pa(A_i)}$ defines a process operator.  (See Methods for an explicit demonstration of this fact.) The constraint in the definition of a QCM that $\sigma_{A_1 ... A_n} = \prod_i \rho_{A_i | Pa(A_i)}$ is a valid process operator is essential, and is what guarantees that grandfather-type paradoxes do not arise \cite{OreshkovEtAl_2012_QuantumCorrelationsWithoutCausalOrder}. This is in contrast to the acyclic case, where, given an acyclic causal structure, it is not hard to argue that any product of commuting channels of the form $\prod_i \rho_{A_i | Pa(A_i)}$ is a valid process operator  \cite{BarrettEtAl_2019_QCMs}, hence in particular a faithful QCM with that causal structure can always be found.

\subsection*{Unitarity and causal structure}

The definition of a QCM above is predicated on the idea that causal structure should be represented by a directed graph. This idea, however, along with the stipulation that the accompanying process is Markov for the graph, was presented without much justification or further comment. Why is causal structure represented by a directed graph, for example, as opposed to a different mathematical object, such as a partial order, or a preorder, or some kind of hypergraph? This section considers a subclass of processes -- unitary processes, defined momentarily -- and shows that a unitary process is associated with a causal structure, which can indeed be represented with a directed graph, and that the unitary process is Markov for that graph. In other words, a unitary process, along with its causal structure, defines a QCM. 

In order to define a unitary process, observe that a process operator $\sigma_{A_1 ... A_n}$ has the mathematical form of the CJ operator for a channel $\mathcal{P}: \mathcal{L}( \bigotimes_i \mathcal{H}_{A^{\text{out}}_i} ) \rightarrow \mathcal{L}( \bigotimes_i \mathcal{H}_{A^{\text{in}}_i} )$ \cite{OreshkovEtAl_2012_QuantumCorrelationsWithoutCausalOrder}. 
Where it is convenient to emphasise this form, we will sometimes write $\sigma_{A_1 ... A_n} = \rho^{\mathcal{P}}_{A_1...A_n|A_1...A_n}$, where it is understood implicitly that an $A_i$ to the right of the bar stands for $A^{\text{out}}_i$, while an $A_i$ to the left of the bar stands for $A^{\text{in}}_i$. 
A unitary process is a process (where some of the input or output spaces may be trivial, i.e., $1$-dimensional) such that the channel $\mathcal{P}$ is a unitary channel. 

The first step is to define a notion of causal structure that pertains to the inputs and outputs of a unitary channel. 

\begin{definition} \textnormal{(Causal structure of a unitary channel)} \label{Def_CausalStructureChannel} 
Given a unitary channel $\rho^{\mathcal{U}}_{CD|AB}$, write $A \nrightarrow D$ ($A$ \textnormal{does not influence} $D$), iff $\Trace_{C} [ \rho^{\mathcal{U}}_{CD|AB} ] = \rho^{\mathcal{M}}_{D|B} \otimes \mathds{1}_{A^*}$ for some marginal channel $\mathcal{M}$. 
If $A$ can influence $D$, i.e. $\neg (A \nrightarrow D)$, $A$ is a \textnormal{direct cause} of $D$. 
For any unitary channel $\rho^{\mathcal{U}}_{C_1...C_l|B_1...B_k}$ with $k$ input and $l$ output subsystems its \textnormal{causal structure} is then the set of causal relations between input and output subsystems and can be represented by a DAG with vertices $B_1,...,B_k$ and $C_1,...,C_l$  and an arrow $B_j \rightarrow C_i$ whenever $B_j$ is a \textnormal{direct cause} of $C_i$. 
\end{definition}

This definition (which, given the correspondence between unitary maps $U$ and unitary channels $\mathcal{U}(\_ ) = U(\_ )U^{\dagger}$, we let refer to either) lifts naturally to the case of a unitary process, in such a way that causal relationships are defined between the nodes of the process, rather than between inputs and outputs of a channel.
\begin{definition} \textnormal{(Causal structure of a unitary process)} \label{Def_UnitaryProcess}
Given a unitary process $\sigma_{A_1 ... A_n}=\rho^{\mathcal{U}}_{A_1...A_1|A_1...A_n}$, write $A_j \nrightarrow A_i$ (node $A_j$ \textnormal{does not influence} node $A_i$), iff $A_j^{\textrm{out}}$ does not influence  $A_i^{\textrm{in}}$ in $\mathcal{U}$. 
If node $A_j$ can influence node $A_i$, then $A_j$ is a \textnormal{direct cause} of $A_i$. The \textnormal{causal structure} of the unitary process is the set of all causal relations between its quantum nodes, and is representable as the directed graph with vertices $A_1, ..., A_n$ and an arrow $A_j \rightarrow A_i$, whenever  $A_j$ is a direct cause of $A_i$.
\end{definition}

The fact that any unitary process is Markov for its causal structure, hence defines a QCM, is then immediate from the following theorem of Refs.~\cite{AllenEtAl_2016_QCM,BarrettEtAl_2019_QCMs}. 
\begin{theorem}	\textnormal{(\cite{AllenEtAl_2016_QCM,BarrettEtAl_2019_QCMs})} \label{Thm_FactorizationUnitary}
Given a unitary channel $\rho^{\mathcal{U}}_{C_1...C_l|B_1...B_k}$, let $\{Pa(C_i)\}_{i=1}^l$ be the parental sets as defined by its causal structure. Then the CJ operator factorizes as $\rho^{\mathcal{U}}_{C_1...C_l|B_1...B_k} = \prod_{i=1}^l \rho_{C_i | Pa(C_i)} $, where the marginal channels commute pairwise, $[\rho_{C_i |  Pa(C_i)} \ , \ \rho_{C_j |  Pa(C_j)} ]=0$ for all $i,j$.
\end{theorem}

The case of non-unitary processes, and their relationship to causal structure is presented below.   
First, we describe a well-known example of a causally nonseparable process -- the quantum SWITCH  \cite{ChiribellaEtAl_2013_QuantumCompWithoutDefCausalStructure} -- and show explicitly that it defines a unitary process operator with cyclic causal structure, hence a cyclic QCM.

\subsection*{Example: The Quantum SWITCH} 

The quantum SWITCH \cite{ChiribellaEtAl_2013_QuantumCompWithoutDefCausalStructure} was the first example described of a causally non-separable process. The SWITCH is standardly defined as a higher-order map \cite{ChirbiellaEtAl_2009_QuantumNetworkFramework, ChiribellaEtAl_2013_QuantumCompWithoutDefCausalStructure, BisioEtAl_2019_TheoreticalFrameworkForHigherOrderQuantumTheory} that takes as input two CP maps 
$\mathcal{F}_A:\mathcal{L}(\mathcal{H}_{A^{\textrm{in}}})\rightarrow \mathcal{L}(\mathcal{H}_{A^{\textrm{out}}})$ and 
$\mathcal{G}_B:\mathcal{L}(\mathcal{H}_{B^{\textrm{in}}})\rightarrow \mathcal{L}(\mathcal{H}_{B^{\textrm{out}}})$, where $d_{A^{\textrm{in}}}= d_{A^{\textrm{out}}}=d_{B^{\textrm{in}}}=d_{B^{\textrm{out}}}=d$, and gives as an output a CP map 
$\mathcal{E}:\mathcal{L}(\mathcal{H}_{Q}\otimes \mathcal{H}_{S}) \rightarrow \mathcal{L}(\mathcal{H}_{Q'}\otimes \mathcal{H}_{S'})$, 
where $d_Q=d_{Q'}=2$ and $d_S=d_{S'}=d$. Here, $\mathcal{H}_Q$ and $\mathcal{H}_{Q'}$ are interpreted as the Hilbert spaces of a control qubit at some initial and some final time, respectively, and $\mathcal{H}_S$ and $\mathcal{H}_{S'}$ as the Hilbert spaces of some target system at the same two times.
Intuitively, the effect of the quantum SWITCH is to transform the target system from the initial to the final time by the sequential application of the CP maps $\mathcal{F}_A$ and $\mathcal{G}_B$, where the order in which the two CP maps are applied is conditioned coherently on the logical value of the control qubit. 

\begin{center}
	\begin{figure}
		\centering
		\vspace*{-0.3cm}
		\hspace*{0.1cm}
		\includegraphics[scale=1.0]{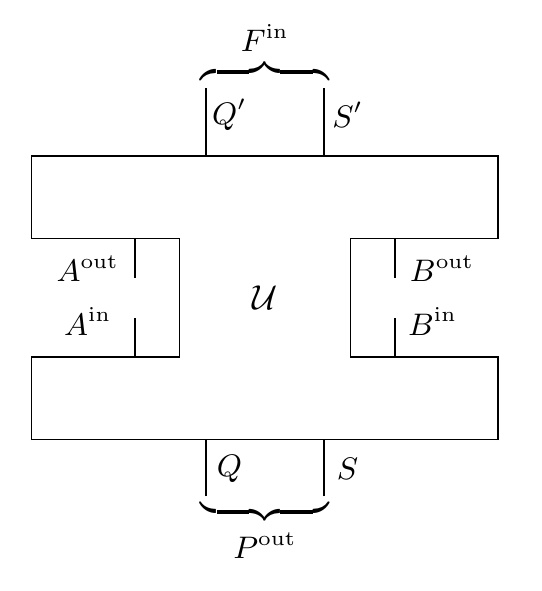}
		\vspace*{-0.3cm}
		\caption{A unitary process with nodes $A$ and $B$, root node $P$ (in the global past) and leaf node $F$ (in the global future). Here with $P^{\text{out}}=QS$ and $F^{\text{in}}=Q' S'$; the quantum SWITCH is an example of such a process.\label{Fig_SWITCH}}
	\end{figure}
\end{center}

\begin{figure}
	\centering
	\vspace*{-0.3cm}
	\hspace*{0.1cm}
	\includegraphics[scale=1.0]{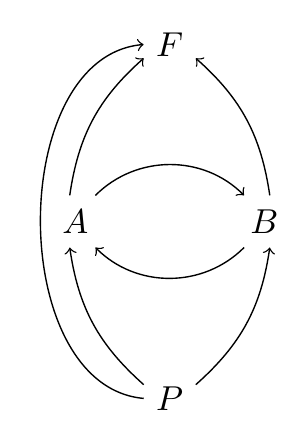}
	\vspace*{-0.3cm}
	\\[0.3cm]
	\begin{minipage}{6cm}
		\centering
		\caption{The causal structure $G_{\tiny{\text{SWITCH}}}$ of the quantum SWITCH.\label{Fig_SWITCH_DG}}
	\end{minipage}
\end{figure}

To formulate this precisely, we will describe the quantum SWITCH directly as a 4-node process (see Fig.~\ref{Fig_SWITCH}), which involves the nodes $A$ and $B$, where $\mathcal{F}_A$ and $\mathcal{G}_B$ are inserted, a node $P$ with $P^{\text{out}}=QS$, where the control qubit and target system at the initial time are prepared in some state, and node $F$ with $F^{\text{in}}=Q' S'$, where the control qubit and the system at the final time are subject to some measurement. 
The SWITCH is then a unitary four-partite process with process operator 
$\sigma^{\textrm{\tiny{SWITCH}}}_{ABPF} = \rho^{\mathcal{U}}_{ABF|ABP} =  \ket{W} \bra{W}$, where
\begin{eqnarray}\label{switchformofw} 
\ket{W} &:=& \ket{0}_{Q^*} \ket{0}_{Q'} \ket{\phi^+}_{S^*A^{\text{in}}} \ket{\phi^+}_{(A^{\text{out})^*}B^{\text{in}}} \ket{\phi^+}_{(B^{\text{out}})^* S'} \nonumber \\
	&&	\hspace{-30pt}  +  \ket{1}_{Q^*} \ket{1}_{Q'} \ket{\phi^+}_{S^*B^{\text{in}}} \ket{\phi^+}_{(B^{\text{out}})^*A^{\text{in}}} \ket{\phi^+}_{(A^{\text{out}})^* S'},  
\end{eqnarray} 
with $\ket{\phi^+}_{XY}:=\sum_i \ket{i}_{X}\ket{i}_{Y}$ and the appearance of the dual spaces due to our convention for the CJ isomorphism. 
It is straightforward to verify that the causal structure of $\sigma^{\textrm{\tiny{SWITCH}}}_{ABPF}$ is the cyclic directed graph in Fig.~\ref{Fig_SWITCH_DG}. 
From Thm.~\ref{Thm_FactorizationUnitary}, it follows that 
\begin{equation}
	\sigma^{\textrm{\tiny{SWITCH}}}_{ABPF} = \rho_{F|ABP} \ \rho_{A|BP} \ \rho_{B|AP} \ \rho_P \ , \label{Eq_SWITCHAsQCM}
\end{equation}
where we have formally added $\rho_P$ to make the Markovianity of $\sigma^{\textrm{\tiny{SWITCH}}}_{ABPF}$ for $G_{\tiny{\text{SWITCH}}}$ explicit, but here $\rho_P$ is just the number $1$, since $P^{\text{in}}$ is trivial. 
Hence, the graph $G_{\tiny{\text{SWITCH}}}$ together with  $\rho_{F|ABP}$, $\rho_{A|BP}$, $ \rho_{B|AP}$, $\rho_P$, form a faithful cyclic QCM. 

\subsection*{Compatibility vs Markovianity}

This section extends the discussion of causal structure to non-unitary processes. 
Briefly, in a QCM involving a non-unitary process $\sigma$, the arrows of the graph are taken to represent facts about the causal structure of some underlying unitary process, with the property that $\sigma$ is recovered from the unitary process when marginalizing over auxiliary systems. The auxiliary systems take the form of a final system $F$, along with uncorrelated local disturbances, where the latter are inputs to the unitary process in a direct product state, with the property that each of them is a direct cause of at most one of the nodes of $\sigma$. As we shall show, it then follows that the process $\sigma$ is Markov for the graph.

The following was introduced in Ref.~\cite{AraujoEtAl_2017_PurificationPostulateForQMWithIndefiniteCausalOrder} 
(there under the name purifiability), 
and will help make these ideas precise.

\begin{definition} \textnormal{(Unitary extendibility)}
	A process $\sigma_{A_1 ... A_n}$ is called \textnormal{unitarily extendible} iff there exists a unitary process $\sigma_{A_1 ... A_n PF} = \rho^{\mathcal{U}}_{A_1 ... A_n F | A_1 ... A_n P}$ on the quantum nodes $A_1 , \ldots , A_n$, plus   additional root node $P$ and leaf node $F$, such that  
	$\sigma_{A_1 ... A_n} = \Trace_{FP} [\sigma_{A_1 ... A_n PF} \ \tau_P]$ 
	for some state $\tau_P \in \mathcal{L}(\mathcal{H}^*_{P^{\text{out}}})$. 	
	The process $\sigma_{A_1 ... A_n PF} $ is called a \textnormal{unitary extension} of $\sigma_{A_1 ... A_n}$.
\end{definition}
It was found in Ref.~\cite{AraujoEtAl_2017_PurificationPostulateForQMWithIndefiniteCausalOrder} that not all process operators are unitarily extendible. The reason for this is that, although for any process $\sigma_{A_1 ... A_n} = \rho^{\mathcal{P}}_{A_1 ... A_n | A_1 ... A_n }$, corresponding to a channel $\mathcal{P}$, the channel $\mathcal{P}$ admits a dilation to a unitary channel, this unitary channel does not necessarily correspond to a valid process itself. Process operators that are not unitarily extendible are those for which no dilation exists such that the unitary channel corresponds to a valid process.

Now suppose that a process $\sigma_{A_1 ... A_n}$ does have a unitary extension $\sigma_{A_1 ... A_n PF}$, involving the additional root node $P$. As per Def.~\ref{Def_UnitaryProcess}, the unitary extension $\sigma_{A_1 ... A_n PF}$ has a causal structure given by some directed graph $G$ with nodes $A_1, ..., A_n , P, F$. Let $G'$ be the subgraph with nodes $A_1 , ... , A_n$, along with all arrows that connect only these nodes in $G$. In general, in the graph $G$, the node $P$ will have arrows to several of the $A_i$, meaning that $P$ is a common cause for these nodes. 
There will then, in general, be correlations in $\sigma$ that are explained by the common cause $P$. This means that the graph $G'$, which omits $P$, is at best an incomplete causal explanation for the correlations in $\sigma$, since it does not explain those correlations due to $P$. 
In this case, there is no reason why $\sigma$ should be Markov for the graph $G'$.

Consider now a unitary extension of $\sigma_{A_1 ... A_n}$ with the feature that the node $P$ can be factored into uncorrelated local disturbances $\lambda_i$, such that each $\lambda_i$ is a direct cause of at most one of the nodes $A_i$. In this case, the graph $G'$, obtained by omitting all of the $\lambda_i$ and leaf node $F$, can be seen as a causal explanation for correlations described by the process $\sigma_{A_1 ... A_n}$, which omits only local disturbances and the final effect $F$, and which does not omit common causes. In this case, we will say that $\sigma$ is compatible with the graph $G'$. In fact, it is more useful to define this term more broadly: we will say that $\sigma$ is compatible with any graph, with nodes $A_1 , ... , A_n$, that contains $G'$ as a subgraph. The following definition makes this precise, generalizing that of Ref.~\cite{BarrettEtAl_2019_QCMs} to the cyclic case. 
\begin{definition} \textnormal{(Compatibility with a directed graph)} \label{Def_Compatibility}
A process $\sigma_{A_1 ... A_n}$ is \textnormal{compatible} with a directed graph $G$ with nodes $A_1, ... ,A_n$, iff $\sigma_{A_1 ... A_n}$ is extendible to a unitary process $\sigma_{A_1 ... A_n \lambda_1...\lambda_n F}$, with an extra root node $\lambda_i$ for $i=1,...,n$ and an extra leaf node $F$, such that:
\begin{enumerate}[label=\textnormal{(\arabic*)}, leftmargin=0.7cm]
	\item there exists a product state $\tau_{\lambda_1} \otimes \cdots \otimes \tau_{\lambda_n}$ with 
	$\tau_{\lambda_i} \in \mathcal{L}(\mathcal{H}^*_{\lambda_i^{\text{out}}})$ such that 
	$\sigma_{A_1 ... A_n}  =  \Trace_{\lambda_1...\lambda_n F} \left[ \ \sigma_{A_1 ... A_n \lambda_1...\lambda_n F} \ ( \tau_{\lambda_1} \otimes \cdots \otimes \tau_{\lambda_n} ) \right] $, 
	\item $\sigma_{A_1 ... A_n \lambda_1...\lambda_n F}$ satisfies the following no-influence conditions 
	(with $Pa(A_i)$ referring to $G$): $\left\{ A_j \nrightarrow A_i \right\}_{A_j \notin Pa(A_i)} \ , \ \left\{ \lambda_j \nrightarrow A_i \right\}_{j \neq i}$. 
\end{enumerate}
\end{definition} 

The following then justifies the stipulation, as a part of the definition of a QCM, that the process accompanying a graph is Markov for the graph.
\begin{theorem} \label{Thm_CompImpliesMarkov}
If a process $\sigma_{A_1 ... A_n}$ is compatible with the directed graph $G$, then it is also Markov for $G$.
\end{theorem}

\noindent \textit{Proof:} 
Similarly to the acyclic case in Ref.~\cite{BarrettEtAl_2019_QCMs}, the theorem follows essentially from Thm.~\ref{Thm_FactorizationUnitary}: the unitary extension, asserted to exist by virtue of the assumed compatibility with $G$, has to factorize into pairwise commuting operators of the form $\sigma_{A_1 ... A_n \lambda_1...\lambda_n F}= \rho_{F|A_1...A_n\lambda_1...\lambda_n} \Big( \prod_i  \rho_{A_i|Pa(A_i) \lambda_i} \Big) $. 
This yields $\sigma_{A_1 \cdots A_n}  =  \prod_i  \Trace_{\lambda_i} \left[ \ \rho_{A_i|Pa(A_i)\lambda_i} \ \tau_{\lambda_i} \ \right] $, where the factors 
$\rho_{A_i|Pa(A_i)}:= \Trace_{\lambda_i} \left[ \ \rho_{A_i|Pa(A_i)\lambda_i} \ \tau_{\lambda_i} \right]$ are pairwise commuting operators.  

Ref.~\cite{BarrettEtAl_2019_QCMs} also establishes a converse to this result, for the case that $G$ is acyclic. For a general directed graph $G$, however, the same proof does not suffice since, even though a dilation to a unitary channel with the required causal constraints can always be found \cite{BarrettEtAl_2019_QCMs}, it is not immediate whether this channel can be guaranteed to define a valid process. We pose this as a hypothesis: 
\begin{hypothesis} \label{quantumconjecture}
If a process $\sigma_{A_1 ... A_n}$ is Markov for a directed graph $G$, then it is compatible with $G$.
\end{hypothesis} 
The hypothesis is satisfied by all examples that we have investigated, but we do not have a proof that it is true in general. Some consequences of the validity or otherwise of this hypothesis are discussed in the Conclusions.

\subsection*{Example: a process that violates a causal inequality}

While the quantum SWITCH is causally nonseparable, the correlations that can be established between operations at the nodes of the quantum SWITCH can always be obtained by a causally separable process with sufficiently large input and output dimensions at each node \cite{AraujoEtAl_2015_WitnessingCausalNonSeparability, OreshkovEtAl_2016_CausallySeparableProcesses}. 
There are, however, causally nonseparable processes that can produce correlations violating causal inequalities \cite{OreshkovEtAl_2012_QuantumCorrelationsWithoutCausalOrder, BaumelerEtAl_2014_PerfectSignallingAmongThreeParties, 
BaumelerEtAl_2014_MaximalIncompatibilityCausalStructure, 
BranciardEtAl_2015_SimplestCausalInequalitiesAndViolations, 
BhattacharyaEtAL_2015_BiasedNonCausalGame, 
FeixEtAl_2016_CausallyNonSeparableProcessesAdmittingCausalModel, 
AbbottEtAl_2016_MultipartiteCausalCorrelations, 
Baumeler_EtAl_2016_SpaceOfLOgicallyConsistentClassicalProcesses, 
OreshkovEtAl_2016_CausallySeparableProcesses}, which are incompatible with the existence of a definite order between the nodes irrespectively of the types of systems or operations performed at those nodes \cite{Chiribella_2012_PerfectDiscriminationOfChannelsViaSuperpositionCS, OreshkovEtAl_2016_CausallySeparableProcesses, Branciard_2016_WitnessesesOfCausalNonseparability}.  
In the literature such processes are called noncausal \cite{OreshkovEtAl_2016_CausallySeparableProcesses}.

\begin{figure}
	\centering
	\vspace*{0.5cm}
	\hspace*{0.1cm}
	\includegraphics[scale=1.0]{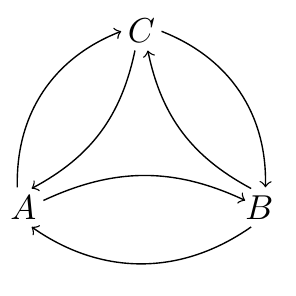}
	\vspace*{-0.3cm}
	\\[0.3cm]
	\begin{minipage}{7cm}
		\centering
		\caption{The causal structure of the AF process.\label{Fig_BWProcess_DG}}
	\end{minipage}
\end{figure}

An example of a tripartite noncausal process is the one which was found by Ara{\'u}jo and Feix (AF) and then published and further studied by Baumeler and Wolf in Ref.~\cite{Baumeler_EtAl_2016_SpaceOfLOgicallyConsistentClassicalProcesses, Baumeler_PhDThesis}. 
It is remarkable in that the process is both classical and deterministic (see  below  for further discussion of classical processes). Any classical process can be viewed as a quantum process, diagonal with respect to a product basis. The AF process, viewed as a quantum process on nodes $A$, $B$ and $C$, each with two-dimensional input and output Hilbert spaces, is described by the process operator
\begin{gather}
\sigma^{\textrm{\tiny{AF}}}_{ABC} = \rho_{A|BC} \ \rho_{B|CA} \ \rho_{C|AB},  
\end{gather}
where 
\begin{gather}
\rho_{A|BC} = \hspace*{-0.25cm} \sum_{b,c = 0,1} |\neg b \wedge c \rangle \langle \neg b \wedge c |_{A^{\textrm{in}}}\otimes |b,c\rangle\langle b,c|_{(B^{\textrm{out}} C^{\textrm{out}})^*},  \\
\rho_{B|CA} = \hspace*{-0.25cm} \sum_{c,a = 0,1} |\neg c \wedge a \rangle \langle \neg c \wedge a |_{B^{\textrm{in}}}\otimes |c,a\rangle\langle c,a|_{(C^{\textrm{out}} A^{\textrm{out}})^*},  \\
\rho_{C|AB} = \hspace*{-0.25cm} \sum_{a,b = 0,1} |\neg a \wedge b \rangle \langle \neg a \wedge b |_{C^{\textrm{in}}}\otimes |a,b\rangle\langle a,b|_{(A^{\textrm{out}} B^{\textrm{out}})^*}. 
\end{gather}
As is explicit in this description, the AF process together with the causal structure in Fig.~\ref{Fig_BWProcess_DG} defines a faithful cyclic QCM.

It was shown by Baumeler and Wolf (BW) \cite{Baumeler_PhDThesis} that 
this process is unitarily extendible (also see Refs.~\cite{AraujoEtAl_2017_QuantumComputationWithIndefiniteCausalStructure, AraujoEtAl_2017_PurificationPostulateForQMWithIndefiniteCausalOrder}) with a unitary extension given by
\begin{eqnarray}
	\sigma^{\textrm{\tiny{BW}}}_{ABCFP} = \rho^U_{ABC F|ABC P} \ , \label{Eq_BWUnitaryExtension}
\end{eqnarray}
where the output space of the root node $P$ is a tensor product of three qubits $\mathcal{H}_{P^{\textrm{out}}} = \mathcal{H}_{\lambda_A} \otimes \mathcal{H}_{\lambda_B} \otimes \mathcal{H}_{\lambda_C}$ and the unitary $U$ is defined by the following bijection of orthonormal bases:
\begin{eqnarray} \label{Eq_DefBWExtensionUNitaryBijection}
	&&U  : \ | a,b,c \rangle_{A^{\textrm{out}} B^{\textrm{out}} C^{\textrm{out}}} 
 \ \otimes \ | l,m,n\rangle_{\lambda_A\lambda_B \lambda_C} \nonumber \\
  && \ \mapsto  \ | l \oplus (\neg b \wedge c) , \ m \oplus (\neg c \wedge a) , \ n \oplus (\neg a \wedge b) \rangle_{A^{\textrm{in}} B^{\textrm{in}} C^{\textrm{in}}}  \nonumber 	\\ 
 && \hspace*{0.7cm} \otimes \ | a,b,c \rangle_{F^{\text{in}}} \ . 
\end{eqnarray}
The original AF process is recovered for marginalization over $F$ and feeding in the product state $| 0,0,0 \rangle$ for $\lambda_A$, $\lambda_B$ and $\lambda_C$. Formally letting the latter three define distinct root nodes $\lambda_A$, $\lambda_B$ and $\lambda_C$, it is not too hard to show that this BW unitary extension also satisfies the corresponding causal constraints of Def.~\ref{Def_Compatibility} to establish $\sigma^{\textrm{\tiny{AF}}}_{ABC}$ to be compatible with the graph of Fig.~\ref{Fig_BWProcess_DG} -- in keeping with Hypothesis~\ref{quantumconjecture}. 
	
\subsection*{Cyclicity and extended circuit diagrams} 

An essential feature of the Markov condition in Def.~\ref{Def_MarkovCondition} is the pairwise commutation relation of the operators of the form $\rho_{A_i|Pa(A_i)}$, where the parental sets in general overlap. 
That two commuting operators act non-trivially on the same Hilbert space has consequences for the algebraic structure of the operators and leads to an intimate link between causal and compositional structure. 
In order to exemplify the fruitfulness of studying this link the following will revisit the two examples from earlier.

\begin{figure}
	\centering
	\vspace*{-0.3cm}
	\hspace*{0.1cm}
	\includegraphics[scale=1.0]{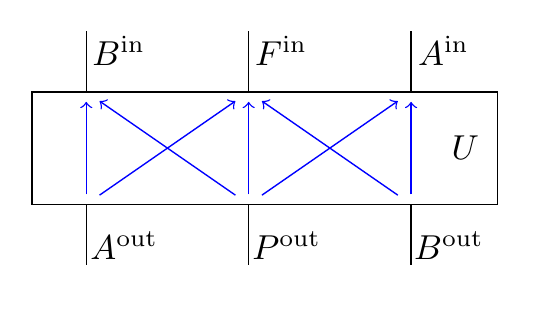}
	\vspace*{-0.6cm}
	\begin{minipage}{8cm} 
		\caption{Unitary map $U$ that defines the quantum SWITCH. The causal structure of $U$ is indicated in blue. \label{Fig_UEUnitaryWithCausalStructure}}
	\end{minipage}
\end{figure}

The quantum SWITCH can be considered as a unitary process over $4$ nodes, given by  
$\sigma^{\textrm{\tiny{SWITCH}}}_{ABPF} = \rho^{\mathcal{U}}_{ABF|ABP} =  \ket{W} \bra{W}$, where $\ket{W}$ is defined in Eq.~(\ref{switchformofw}). The unitary channel $\mathcal{U}$ corresponds to a unitary map $U: \mathcal{H}_{A^{\textrm{out}}} \otimes  \mathcal{H}_{P^{\textrm{out}}} \otimes  \mathcal{H}_{B^{\textrm{out}}}  \rightarrow \mathcal{H}_{A^{\textrm{in}}} \otimes  \mathcal{H}_{F^{\textrm{in}}} \otimes  \mathcal{H}_{B^{\textrm{in}}}$,  
which is depicted in Fig.~\ref{Fig_UEUnitaryWithCausalStructure} together with its causal structure shown in blue. Observe in particular that in $U$, $A^{\textrm{out}}$ does not influence $A^{\textrm{in}}$, and similarly $B^{\textrm{out}}$ does not influence $B^{\textrm{in}}$, as must be the case for a well-defined process \cite{Oreshkov_2019_TimeDelocalizedSubsystems}. 

Ref.~\cite{LorenzEtAl_2020_CausalAndCompStructure} shows that any unitary map $U$ with three in- and output systems, and the causal constraints of Fig.~\ref{Fig_UEUnitaryWithCausalStructure}, has a decomposition of the following form: 
\begin{eqnarray}
	U = \Big(\mathds{1}_{B^{\textrm{in} } } \otimes T \otimes \mathds{1}_{A^{\textrm{in} } } \Big) \Big( \bigoplus_{i \in I} V_i \otimes W_i \Big) \ \hspace*{2.2cm} \nonumber \\[-0.3cm]
	 \hspace*{1.3cm} \Big(\mathds{1}_{A^{\textrm{out} } }  \otimes S \otimes \mathds{1}_{B^{\textrm{out}} } \Big),  \hspace*{0.6cm} \label{subsystemdecomposition}
\end{eqnarray}
where 
$S$ and $T$ are unitaries, and $\{V_i\}_{i\in I}$ and $\{W_i\}_{i\in I}$ families of unitaries of the form
\begin{eqnarray}
S &:&  \ \mathcal{H}_{P^{\textrm{out}}} \ \rightarrow \ \bigoplus_{i\in I} \mathcal{H}_{P_i^L} \otimes \mathcal{H}_{P_i^R} \ ,  \\
V_i &:& \ \mathcal{H}_{A^{ \textrm{out}}} \otimes \mathcal{H}_{P_i^L} \ \rightarrow \ \mathcal{H}_{B^{\textrm{in} } } \otimes \mathcal{H}_{F_i^L} \ ,  \\
W_i &:& \ \mathcal{H}_{P_i^R} \otimes \mathcal{H}_{B^{\textrm{out} }  } \ \rightarrow \ \mathcal{H}_{F_i^R} \otimes \mathcal{H}_{A^{\textrm{in} } } \ ,  \\
T &:& \  \bigoplus_{i\in I} \mathcal{H}_{F_i^L} \otimes \mathcal{H}_{F_i^R} \ \rightarrow \ \mathcal{H}_{F^{\textrm{in}}} \ . 
\end{eqnarray}

\begin{figure}
	\centering
	\vspace*{-0.7cm}  
	\includegraphics[scale=1.0]{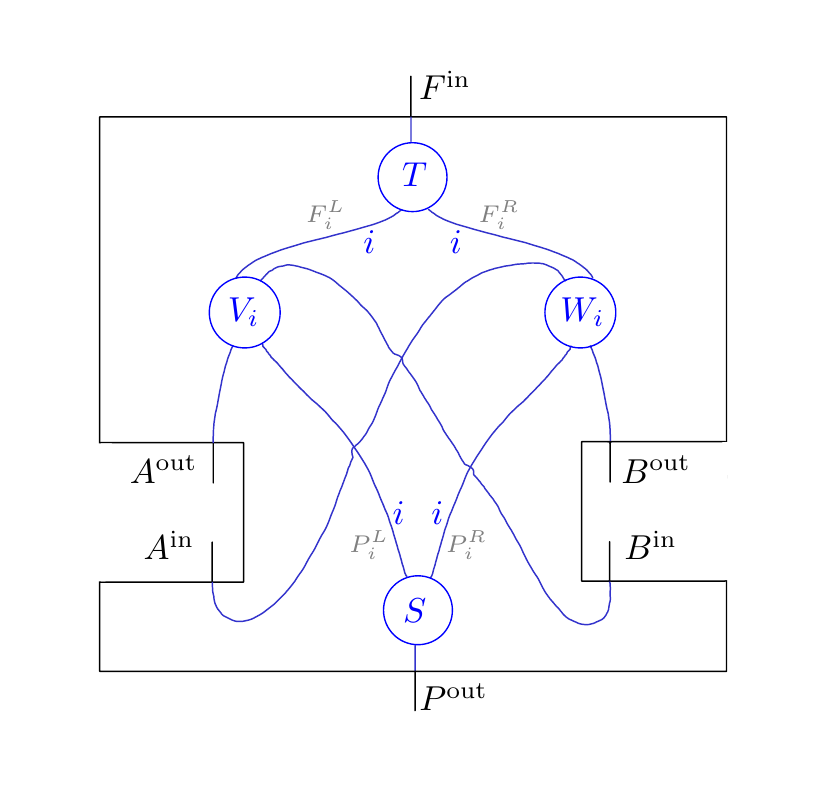}
	\begin{minipage}{8cm} 
	\vspace*{-0.4cm}
	\caption{Extended circuit diagram decomposition of the quantum SWITCH. Additionally indicated in gray are the labels of the intermediate families of Hilbert spaces. \label{SWITCHExtendedCircuitDecomposition}}
	\end{minipage}
\end{figure}

Such a compositional structure with direct sums over tensor products goes beyond what is expressible with ordinary circuit diagrams. Ref.~\cite{LorenzEtAl_2020_CausalAndCompStructure} therefore introduced extended circuit diagrams to give a graphical representation of such decompositions. 
Fig.~\ref{SWITCHExtendedCircuitDecomposition} arises from that extended circuit diagram representation of Eq.~\ref{subsystemdecomposition} by bending the wires corresponding to $A^{\textrm{in}}$ and $B^{\textrm{in}}$ down to re-identify the quantum nodes $A$ and $B$ -- thereby filling the black box of the quantum SWITCH from Fig.~\ref{Fig_SWITCH}.  
For details on this diagrammatic language we refer the reader to Ref.~\cite{LorenzEtAl_2020_CausalAndCompStructure}, but the essential idea is that individual wires with indices on them, such as those between the circles $S$ and $V_i$ and $W_i$, respectively, represent the families of Hilbert spaces $\{ \mathcal{H}_{P_i^L} \}_{i \in I}$ and $\{ \mathcal{H}_{P_i^R} \}_{i \in I}$, while the two parallel wires together represent $\bigoplus_{i\in I} \mathcal{H}_{P_i^L} \otimes \mathcal{H}_{P_i^R}$. An implicit summation over orthogonal subspaces indexed by $i$ allows the representation of the intermediate unitary map $\bigoplus_i V_i \otimes W_i$ from Eq.~\ref{subsystemdecomposition}. 

It is easy to see what this decomposition is concretely in the case of the quantum SWITCH: the index $i$ takes two values, $0$ and $1$, corresponding to the logical values of the control qubit, i.e., $\mathcal{H}_P = \mathcal{H}_Q \otimes \mathcal{H}_S \cong (\mathds{C} \otimes \mathcal{H}_S ) \oplus (\mathcal{H}_S \otimes \mathds{C})$ and the unitaries $V_i$ and $W_i$ are either the SWAP transformation on the respective systems or the identity depending on $i$. We see that even though the causal structure of the full process is cyclic, the process splits into a direct sum of processes in each of which causal influence and the flow of information follow acyclic paths. 

This decomposition of the quantum SWITCH applies more generally: seeing as any unitary process of the type depicted in Fig.~\ref{Fig_SWITCH}, with a root node $P$, a leaf node $F$, and two nodes $A$ and $B$ in between, satisfies $A^{\textrm{out}} \nrightarrow A^{\textrm{in}}$ and $B^{\textrm{out}} \nrightarrow B^{\textrm{in}}$, it follows that any such unitary process has a decomposition as in Fig.~\ref{SWITCHExtendedCircuitDecomposition}.  
Note that the below will furthermore establish (as a direct consequence of the proof of Thm.~\ref{Thm_ResultCausalSep}) that for each $i$ the summand $V_i \otimes W_i$ of that corresponding decomposition has to have an acyclic causal structure, that is, any unitary process with nodes $A,B,P,F$ where $P$ is a root node and $F$ a leaf node, is a direct sum of unitary processes in which causal influences flow along acyclic paths.

\begin{figure}
	\centering
	\vspace*{-0.3cm}
	\includegraphics[scale=1.0]{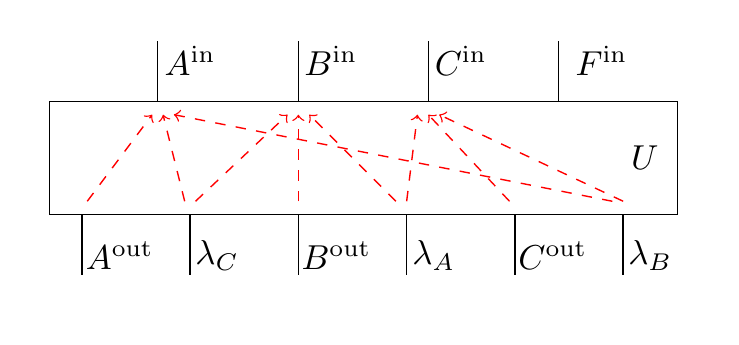}
	\begin{minipage}{8cm} 
	\vspace*{-0.2cm}
		\caption{The unitary map $U$ from Eq.~\eqref{Eq_DefBWExtensionUNitaryBijection} that defines the BW unitary extension of the AF process. Also depicted is its causal structure, where for better visibility, rather than direct cause relations, the no-influence conditions are shown as red dashed arrows. \label{Fig_AFProcessUnitaryExtension}}
	\end{minipage}
\end{figure}

\begin{figure}
	\centering
	\vspace*{-0.3cm}
	\includegraphics[scale=1.0]{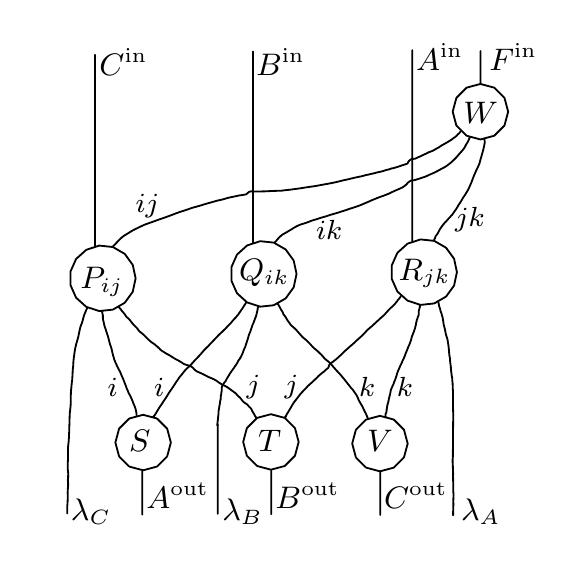}
	\vspace*{-0.3cm}
	\begin{minipage}{8cm} 
		\caption{Causally faithful extended circuit decomposition of the unitary $U$ from Fig.~\ref{Fig_AFProcessUnitaryExtension} (up to some swaps for better readability). \label{Fig_BWExtensionDotDiagramU}}
	\end{minipage}
\end{figure}

The second example concerns the tripartite AF process and its BW unitary extension $\rho^U_{ABC F|ABC P}$ (see Eqs.~\eqref{Eq_BWUnitaryExtension}-\eqref{Eq_DefBWExtensionUNitaryBijection}). 
The root node $P$ has as output space $\mathcal{H}_{P^{\textrm{out}}} = \mathcal{H}_{\lambda_A} \otimes \mathcal{H}_{\lambda_B} \otimes \mathcal{H}_{\lambda_C}$, where each $\lambda_X$ influences only $X$ and $F$ for $X=A,B,C$. 
The associated unitary map $U$ and its causal structure are depicted in Fig.~\ref{Fig_AFProcessUnitaryExtension}. 
The results from Ref.~\cite{LorenzEtAl_2020_CausalAndCompStructure} allow again the statement of an extended circuit decomposition of $U$, which is implied by its causal structure and which makes the pathways of causal influence through $U$ graphically evident (the proof is completely analogous to that of Thm.~7 in Ref.~\cite{LorenzEtAl_2020_CausalAndCompStructure}). 
This decomposition of $U$ is depicted in Fig.~\ref{Fig_BWExtensionDotDiagramU} and reads: 
\begin{eqnarray}
	U &=& \Big( \mathds{1}_{C^{\text{in}}B^{\text{in}}A^{\text{in}}} \otimes W \Big) \
		\Big(\bigoplus_{i,j,k} P_{ij} \otimes Q_{ik} \otimes R_{jk} \Big) \nonumber \\[-0.2cm] 
		&& \Big(  \mathds{1}_{\lambda_C} \otimes S \otimes \mathds{1}_{\lambda_B} \otimes T \otimes V \otimes \mathds{1}_{\lambda_A} \Big) \ , \label{Eq_BWExtensionsDecompositionU}
\end{eqnarray} 
for (families of) unitary maps
\begin{eqnarray}
	S &:& \ \mathcal{H}_{A^{\text{out}}} \ \rightarrow \ \bigoplus_i  \mathcal{H}_{X_i^L} \otimes \mathcal{H}_{X_i^R} \ ,  \\
	T &:& \ \mathcal{H}_{B^{\text{out}}} \ \rightarrow \ \bigoplus_{j}  \mathcal{H}_{Y_j^L} \otimes \mathcal{H}_{Y_j^R} \ ,  \\
	V &:& \ \mathcal{H}_{C^{\text{out}}} \ \rightarrow \ \bigoplus_{k}   \mathcal{H}_{Z_k^L} \otimes \mathcal{H}_{Z_k^R} \ ,  \\
	W &:& \ \bigoplus_{i,j,k}  \mathcal{H}_{G_{ij}^{(1)}} \otimes  \mathcal{H}_{G_{ik}^{(2)}} \otimes  \mathcal{H} _{G_{jk}^{(3)}} \ \rightarrow \ \mathcal{H}_{F^{\text{in}}} \ ,  \\
	P_{ij} &:& \ \mathcal{H}_{\lambda_C} \otimes \mathcal{H}_{X_i^L} \otimes \mathcal{H}_{Y_j^L} \ \rightarrow \ \mathcal{H}_{C^{\text{in}}}  \otimes  \mathcal{H}_{G_{ij}^{(1)}} \ ,	 \\
	Q_{ik} &:& \ \mathcal{H}_{X_i^R} \otimes \mathcal{H}_{\lambda_B} \otimes \mathcal{H}_{Z_k^L} \ \rightarrow \ \mathcal{H}_{B^{\text{in}}}  \otimes  \mathcal{H}_{G_{ik}^{(2)}} \ ,	 \\
	R_{jk} &:& \ \mathcal{H}_{Y_j^R} \otimes \mathcal{H}_{Z_k^R} \otimes \mathcal{H}_{\lambda_A} \ \rightarrow \ \mathcal{H}_{A^{\text{in}}}  \otimes  \mathcal{H}_{G_{jk}^{(3)}} 	\ .  
\end{eqnarray}

\begin{figure}
	\centering
	\vspace*{-0.3cm}
	\hspace*{-0.25cm}
	\includegraphics[scale=1.0]{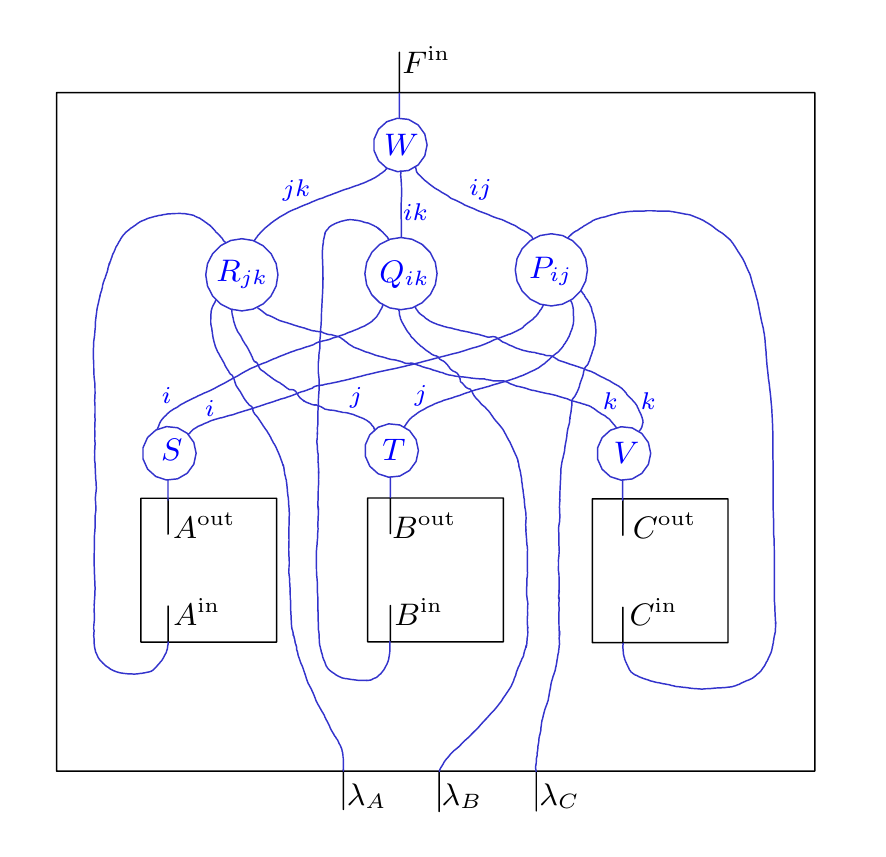}
	\vspace*{-0.3cm}
	\begin{minipage}{8cm} 
		\caption{Extended circuit diagram decomposition of the BW unitary extension of the AF process. \label{Fig_BWExtensionDotDiagramUProcess}}
	\end{minipage}
\end{figure}

By appropriately bending the wires that correspond to $A^{\text{in}}$, $B^{\text{in}}$ and $C^{\text{in}}$ to re-identify the nodes $A$, $B$ and $C$ (and swapping some wires for better readability) one obtains Fig.~\ref{Fig_BWExtensionDotDiagramUProcess}, revealing a fine-grained compositional structure of the BW unitary extension.  

Note that the stated decomposition is general in the sense that a decomposition of the form as in Fig.~\ref{Fig_BWExtensionDotDiagramU} exists for any unitary with a causal structure as in Fig.~\ref{Fig_AFProcessUnitaryExtension}. 
However, in the concrete case of the BW unitary extension one can easily see what the components in Eq.~\eqref{Eq_BWExtensionsDecompositionU} correspond to through a comparison with Eq.~\eqref{Eq_DefBWExtensionUNitaryBijection}. 
All three indices $i$, $j$ and $k$ are binary and, via the unitaries $S$, $T$ and $V$ can be seen to correspond to one-dimensional subspaces of $A^{\text{out}}$ , $B^{\text{out}}$ and $C^{\text{out}}$. 
Hence, each indexed space, i.e. each element of a family of Hilbert spaces associated with an indexed wire, is a trivial Hilbert space. 
For any fixed value $(i,j,k)$, the unitary 
$P_{ij} \otimes Q_{ik} \otimes R_{jk}$ is of the type 
$\mathcal{H}_{\lambda_C} \otimes  \mathcal{H}_{\lambda_B} \otimes \mathcal{H}_{\lambda_A}  \rightarrow  \mathcal{H}_{C^{\text{in}}} \otimes \mathcal{H}_{B^{\text{in}}} \otimes \mathcal{H}_{A^{\text{in}}}$, 
where all spaces are qubits (suppressing all trivial spaces). 
The unitary 
$P_{ij} : \mathcal{H}_{\lambda_C} \rightarrow  \mathcal{H}_{C^{\text{in}}}$ 
maps $|\lambda_C \rangle \mapsto |\lambda_C \oplus (\neg i \wedge j) \rangle$, i.e. $P_{ij}$ is the identity or the NOT gate depending on the values of $i$ and $j$. 
The unitaries $Q_{ik}$ and $R_{jk}$ can similarly be identified through comparison with Eq.~\eqref{Eq_DefBWExtensionUNitaryBijection}.

One thus finds that the BW unitary extension is a direct sum over unitary processes each of which has an acyclic causal structure. 
Furthermore, it is natural to wonder whether knowing a decomposition of the form as in Fig.~\ref{Fig_AFProcessUnitaryExtension} might suggest a way in which the process could be implemented -- a process, which we recall is one that violates a causal inequality.

How about other unitary processes not of the two presented types? 
Ref.~\cite{LorenzEtAl_2020_CausalAndCompStructure} provides extended circuit decompositions for many classes of unitary transformations, where the decompositions are causally faithful, 
meaning that if $A$ is an input to the unitary $U$ and $B$ an output, then there is a path from $A$ to $B$ in the extended circuit iff $A$ can influence $B$ through $U$ (note this is distinct from the notion of faithfulness of a QCM). 
Consider now a unitary map $U$ that corresponds to a unitary process, in the sense that the output Hilbert spaces of the nodes correspond to the inputs to $U$, and the input Hilbert spaces of the nodes correspond to the outputs of $U$. If $U$ has a causally faithful extended circuit decomposition, then by appropriately bending the wires, as in the above examples, one can always obtain a fine-grained compositional structure of the corresponding unitary process. 
Ref.~\cite{LorenzEtAl_2020_CausalAndCompStructure} states the hypothesis that all finite-dimensional unitary transformations (over specified tensor products of input Hilbert spaces and output Hilbert spaces) have a causally faithful extended circuit decomposition. This would mean that all unitary processes, by bending the wires, would admit causally faithful decompositions in a similar manner. At the time of writing, however, the hypothesis remains unproven.

\subsection*{The bipartite unitarily extendible processes}

Understanding which processes have a physical realization is a central open question in the field of indefinite causal order \cite{AraujoEtAl_2017_PurificationPostulateForQMWithIndefiniteCausalOrder, Oreshkov_2019_TimeDelocalizedSubsystems}. While causally nonseparable processes may have a realization in exotic scenarios involving both quantum systems and gravity, it seems clear that any present-day laboratory experiment admits a description in terms of a straightforward, definite, causal ordering of suitably defined parts of the experiment. Nevertheless, various experiments have been performed that are claimed as realizations of nonseparable processes such as the quantum SWITCH \cite{ProcopioEtAl_2015_ExperimentalSuperpositionOfOrders, RubinoEtAl_2017_ExperimentalVerificationIndefiniteCausalOrders, GoswamiEtAl_2018_IndefiniteCausalOrderInQuantumSWITCH, RubinoEtAl_2019_ExperimentalEntanglementOfTemporalOrders, WeiEtAl_2019_ExperimentalQuantumSWITCHCommunicationComplexity, TaddeiEtAl_2020_ExperimentalComputationalAdvantageSuperpositionTemporalOrders}. This has caused some debate \cite{MacleanEtAl_2017_QuantumCoherentMixturesOfCausalRelations, Oreshkov_2019_TimeDelocalizedSubsystems, PaunkovicEtAl_2020_CausalOrdersCircuitsAndSpacetime}. 

Behind much of this debate, however, lies merely a question of how the abstract mathematical description is assumed to map to physical phenomena. Each of the implementations claimed so far is of a process that involves coherent control over the time-ordering of nodes in a similar manner to the SWITCH, and which cannot therefore violate causal inequalities. Ref.~\cite{Oreshkov_2019_TimeDelocalizedSubsystems} shows that any such implementation can be seen as a valid implementation of a nonseparable process, if the process is understood as being defined over time-delocalized systems, where the input and output Hilbert spaces of the nodes of the process correspond to subsystems of tensor products of Hilbert spaces of systems associated with different times. This raises the question: which processes in general admit a laboratory implementation, at least in terms of time-delocalized systems? In particular, can a process violating causal inequalities be implemented?

There was some hope that a process violating causal inequalities could be implemented, because Ref.~\cite{Oreshkov_2019_TimeDelocalizedSubsystems} also shows that every unitary extension of a bipartite process has a realization in terms of time-delocalized systems. Hence if there were a unitarily extendible bipartite process violating causal inequalities, then it could be implemented, at least via time-delocalized systems.  
The following theorem, however, shows that there is no such possibility. Any bipartite unitarily extendible process is causally separable, hence in particular cannot violate causal inequalities, as conjectured in Ref.~\cite{AraujoEtAl_2017_PurificationPostulateForQMWithIndefiniteCausalOrder}; furthermore, all unitary extensions of bipartite processes are variations of the quantum SWITCH, realizable by coherent control of the times of the operations of $A$ and $B$. 
The argument uses the existence of a faithful extended circuit decomposition of the form as in Eq.~\eqref{subsystemdecomposition} that is implied by the causal constraints of Fig.~\ref{Fig_UEUnitaryWithCausalStructure}.  
\begin{theorem} \label{Thm_ResultCausalSep}
All unitarily extendible bipartite processes are causally separable. Given a bipartite process, if it is unitarily extendible, then the unitary extension has a realization in terms of coherent control of the order of the node operations.
\end{theorem}

\noindent \textit{Proof:} See Methods. 

As one can see, e.g., from the AF process, being unitarily extendible does not imply causal separability in the general multipartite case. 
However, the decomposition from Fig.~\ref{Fig_AFProcessUnitaryExtension} of the BW unitary extension of the AF process proved insightful with regards to how the cyclicity of the causal structure comes from different contributions across  the direct sum. 
More generally, suppose a causally faithful extended circuit decomposition of the unitary extension of some multipartite  process is known. 
It is then natural to ask whether some kind of generalization of the constraints established as part of the proof of Thm.~\ref{Thm_ResultCausalSep} could be derived, which in the bipartite case just happen to give causal separability, while in the general case constrain each summand of the decomposition. 
As is the case with the bipartite processes, to which Thm.~\ref{Thm_ResultCausalSep} applies, one would expect that such constraints on summands of the unitary extension also manifest themselves in interesting ways for the non-unitary marginal process.  
We leave this question for future investigation.

\subsection*{Causal nonseparability}

The definition of causal separability was given above only for bipartite processes and it was mentioned that the multipartite case, with more than two nodes, is more intricate.  
This section will first give the general definition, following Refs.~\cite{OreshkovEtAl_2016_CausallySeparableProcesses, WechsEtAl_2019_MultipartiteCausalNonSeparability}, and then present another main result. 

Seeing as the idea of causal separability is to capture whether a process is consistent with our intuitions on causal order, it is natural to let it incorporate the following two features. 
First, in addition to probabilistic mixtures of fixed orders of nodes it allows for a dynamical causal order of them, that is, the overall causal order of some nodes need not be fixed, but may depend on what happens at some earlier nodes. 
Second, it demands that causal separability is preserved under extending the process with an arbitrary ancillary input state shared between the nodes (a property called extensibility \cite{OreshkovEtAl_2016_CausallySeparableProcesses}). 
A process thus is causally separable essentially if, upon considering arbitrary shared entanglement between auxiliary input systems to all nodes, the such extended process can be seen to arise from a probabilistic mixture of particular processes: for each there is a node $P$ in the past such that for all possible interventions at $P$ the marginal process has a fixed causal order, or more generally, is itself again causally separable. 
Hence, one ends up with an iterative definition of the concept. 
This notion was originally called extensible causal separability in Ref.~\cite{OreshkovEtAl_2016_CausallySeparableProcesses} to distinguish it from the analogous concept without extensibility, but as it is undoubtedly the more natural concept, we here refer to it simply as causal separability, as in Ref.~\cite{WechsEtAl_2019_MultipartiteCausalNonSeparability}. 
(Note, there have been two equivalent definitions of that notion \cite{WechsEtAl_2019_MultipartiteCausalNonSeparability}, which differ by whether extensibility is imposed at the level of the full process \cite{OreshkovEtAl_2016_CausallySeparableProcesses} or at each level of the iteration \cite{WechsEtAl_2019_MultipartiteCausalNonSeparability}. For the present purposes, it is convenient to use the latter one.) Finally, making the concept precise relies on the following notion of no-signalling in a process, which, along with various equivalent statements, was given in Ref.~\cite{OreshkovEtAl_2016_CausallySeparableProcesses}.

\begin{definition} \textnormal{(No signalling in a process) \label{Def_NoSigInProcess}} 
Given a process $\sigma_{A_1...A_n}$, we say that there is \textnormal{no signalling} from a subset $S \subset \{A_1, \ldots, A_n\}$ of its nodes to the complementary subset $\overline{S} := \{A_1, \ldots, A_n\} \setminus S $, iff the probabilities $P(k_{\overline{S}})=\Trace \left[\sigma_{A_1...A_n} \left( \tau_{\overline{S}}^{k_{\overline{S}}} \otimes \tau_{S} \right) \right]$ for the outcomes of any operation $ \tau_{\overline{S}}^{k_{\overline{S}}} = \bigotimes_{A\in \overline{S}}  \tau_A^{k_{A}} $ performed at $\overline{S}$ are independent of the choice of trace-preserving operations  $ \tau_{S} = \bigotimes_{A\in S}  \tau_A $ performed at $S$. 
\end{definition}

Now let $\tau_{A_j}$ represent a CP map at the node $A_j$, which is not necessarily trace-preserving. If there is no signalling to a node $A_j$ from $\{A_1, \ldots, A_n\} \setminus \{A_j\}$, then for any $\tau_{A_j}$, the object $\Trace_{A_j} [\sigma_{A_1...A_n} \tau_{A_j} ]$ is proportional to a process operator. In this case, let $\sigma|_{\tau_{A_j}}$ be the corresponding correctly normalized process operator. We refer to $\sigma|_{\tau_{A_j}}$ as a conditional process. 
We can now state the formal definition of causal separability.  

\begin{definition} \textnormal{(Causal separability \cite{WechsEtAl_2019_MultipartiteCausalNonSeparability})} \label{Def_CausalSep}
Every single-node process is \textnormal{causally separable}. For $n\geq 2$, a process $\sigma$ on $n$ quantum nodes $A_1$, $\ldots$, $A_n$ is said to be \textnormal{causally separable}, iff, for any extension of each  node $A_j$ with an additional input system $\mathcal{H}_{(A'_j)^{\text{in}}}$ to a new node $\widetilde{A}_j$, defined by  
	$\mathcal{H}_{\widetilde{A}_j^{\text{in}}} := \mathcal{H}_{A_j^{\text{in}}} \otimes \mathcal{H}_{(A'_j)^{\text{in}}}$ and $\mathcal{H}_{\widetilde{A}_j^{\text{out}}} := \mathcal{H}_{A_j^{\text{out}}}$,
	 and any auxiliary quantum state $\rho \in \mathcal{L}( \mathcal{H}_{(A'_1){\text{in}}} \otimes \ldots \otimes  \mathcal{H}_{(A'_n)^{\text{in}}})$, 
	 the process $\sigma \otimes \rho$ on the quantum nodes $\widetilde{A}_1$, $\ldots$, $\widetilde{A}_n$ decomposes as 
\begin{gather}
\sigma \otimes \rho \ = \ \sum_{k=1}^n q_k \ \sigma^{\rho}_{(k)}, \label{Eq_CSdecomposition} 
\end{gather}
with $q_k\geq 0$, $\sum_k q_k=1$, where for each $k$, $\sigma^{\rho}_{(k)}$ is a process in which there can be no signalling to $\widetilde{A}_k$ from the rest of the nodes, and where for any CP map $\tau_{\widetilde{A}_k}$ that can take place at the node $\widetilde{A}_k$, the conditional process on the remaining $n-1$ nodes, $\sigma^{\rho}_{(k)}|_{\tau_{\widetilde{A}_k} }$, is itself causally separable.
\end{definition}

An important question then concerns the relation between causal nonseparability and cyclicity of causal structure. For a QCM that involves a generic (not necessarily unitary) process, the cyclicity of its directed graph does not in general imply causal nonseparability of the process, even if the QCM is faithful. 
Consider, for example, the quantum SWITCH with process operator $\sigma^{\textrm{\tiny{SWITCH}}}_{ABFP}$. 
Tracing out the system $F^{\textrm{in}}$, we obtain a reduced 3-node process that (relabeling $C$ as $P$) is both faithful and Markov for the graph of Fig.~2b, having the form $\sigma_{ABP} = \rho_{A|BP} \ \rho_{B|AP} \ \rho_P$. 
This process is causally separable, since it can be understood as describing a situation in which the order between $A$ and $B$ depends in an incoherent manner on the logical value of the control qubit prepared at the initial time. 
This process thus forms a faithful cyclic QCM and is a canonical example of a process with dynamical causal order (here between nodes $A$ and $B$). 

In fact, one and the same cyclic graph may appear in two distinct faithful QCMs, one involving a causally separable, the other a nonseparable process. An example of this can again be given using the quantum SWITCH. The latter is causally nonseparable and has the graph in Fig.~\ref{Fig_SWITCH_DG} as causal structure, which however also is the causal structure of the  classical SWITCH \cite{ChiribellaEtAl_2013_QuantumCompWithoutDefCausalStructure}, which in contrast is causally separable (see subsequent discussion of classical processes). 
What this points at is a well-known fact, namely that causal separability cannot separate the distinction between cyclicity and acyclicity on one hand, and classical and quantum causal order on the other hand. 

For the case of unitary processes things are, however, much simpler.

\begin{theorem} \label{Thm_CyclicityAndNonSep}
A unitary process is causally nonseparable iff it has a cyclic causal structure.
\end{theorem}

\noindent \textit{Proof:} See  Methods. 

If a unitary process has a causal structure given by an acyclic graph, then it is a unitary comb \cite{ChirbiellaEtAl_2009_QuantumNetworkFramework}. Hence a unitary process is either a comb or is causally nonseparable -- intermediate possibilities, such as dynamical causal order, cannot arise.  
Note that there is no classical analogue of Thm.~\ref{Thm_CyclicityAndNonSep}, i.e. a classical deterministic process is not necessarily causally nonseparable if it has a cyclic causal structure. 
The classical SWITCH \cite{ChiribellaEtAl_2013_QuantumCompWithoutDefCausalStructure} is again an example that establishes this claim. (See below for an introduction of classical deterministic processes).

\subsection*{Cyclicity and classical processes} 

If a process operator is diagonal in a basis that is a product of local bases for the input and output Hilbert spaces at each node, it is equivalent to a classical process \cite{OreshkovEtAl_2012_QuantumCorrelationsWithoutCausalOrder, BaumelerEtAl_2014_MaximalIncompatibilityCausalStructure, Baumeler_EtAl_2016_SpaceOfLOgicallyConsistentClassicalProcesses}, where each node $X$ is associated with a pair of classical variables $X^{\textrm{in}}$ and $X^{\textrm{out}}$. Following Ref.~\cite{BarrettEtAl_2019_QCMs} we call such classical nodes classical split nodes. Classical processes are studied in detail in Refs.~\cite{BaumelerEtAl_2014_MaximalIncompatibilityCausalStructure, Baumeler_EtAl_2016_SpaceOfLOgicallyConsistentClassicalProcesses, BarrettEtAl_2019_QCMs}. (See also Refs.~\cite{Baumeler_2017_ReversibleTimeTravelWithFreedomOfChoice, TobarEtAl_2020_ReversibleDynamicsWithCTCs}.) This section presents the main ideas, and defines (possibly cyclic) classical split-node causal models. For the most part the definitions are the obvious classical analogues of those for the quantum case. While cyclic classical causal models have sometimes been studied (see, e.g., Refs.~\cite{Richardson_1997_CharacterizationMarkovEquivalenceForDirectedCyclicGraphs,  ForreEtAl_2019_CausalCalculusWithCycles}), for example to encompass the possibility of classical feedback loops, they are not of the split-node variety described here, and are not equivalent. 

A classical process, defined over classical split-nodes $X_1 , ... , X_n$, corresponds to a map $\kappa_{X_1 ... X_n} : X_1^{\text{in}}\times X_1^{\text{out}} \times \cdots\times X_n^{\text{in}}\times X_n^{\text{out}} \rightarrow [0,1]$, such that $\sum_{X_1^{\text{in}}, X_1^{\text{out}},...,X_n^{\text{in}}, X_n^{\text{out}}} \left(\kappa_{X_1 ... X_n} \prod_i P(X_i^{\text{out}} | X_i^{\text{in}})\right) = 1$, for any set of classical channels $\{ P(X_i^{\text{out}} | X_i^{\text{in}})\}$. A local intervention at a node $X$, with outcome $k_X$, corresponds to a classical instrument $P(k_X,X^{\textrm{out}}|  X^{\textrm{in}} )$. Given a local intervention at each node, the joint probability distribution over the outcomes is 
\begin{eqnarray}
P(k_{X_1} , ... , k_{X_n}) &=& \nonumber \\
&& \hspace{-3.5cm} \sum_{X_1^{\text{in}}, X_1^{\text{out}},...,X_n^{\text{in}}, X_n^{\text{out}}} \Big(\kappa_{X_1 ... X_n} \prod_i P(k_{X_i} X_i^{\text{out}} | X_i^{\text{in}})\Big).
\end{eqnarray}

A special case of a classical process is a deterministic process $\kappa_{X_1 ... X_n}^f$, for which $P(X_1^{\text{in}},...,X_n^{\text{in}} | X_1^{\text{out}},...,X_n^{\text{out}}) = \delta( (X_1^{\text{in}} , ... , X_n^{\text{in}}) , \allowbreak f(X_1^{\text{out}} , ... ,  X_n^{\text{out}}))$, where $f : X_1^{\text{out}} \times .... \times X_n^{\text{out}} \allowbreak \rightarrow X_1^{\text{in}} \times .... \times X_n^{\text{in}}$ is a function. When $f$ is bijective, we call such a process reversible. It was shown in Ref.~\cite{Baumeler_EtAl_2016_SpaceOfLOgicallyConsistentClassicalProcesses} that the set of classical processes over nodes $X_1 , ... , X_n$ forms a polytope, and that the deterministic polytope, defined as all convex mixtures of deterministic processes, is in general a strict subset of it. While all classical processes on two nodes are causally separable \cite{OreshkovEtAl_2012_QuantumCorrelationsWithoutCausalOrder}, on three or more nodes there exist classical processes, including deterministic classical processes, that are causally nonseparable -- the AF process from Ref.~\cite{BaumelerEtAl_2014_MaximalIncompatibilityCausalStructure}, described above, is an example. 

\begin{definition} \textnormal{(Classical split-node causal model (CSM) --- generalized)} \label{Def_CSM}
	A \textnormal{CSM} is given by:
\begin{enumerate}[label=\textnormal{(\arabic*)}, leftmargin=0.7cm]
\item a causal structure represented by a directed graph $G$ with vertices corresponding to classical split-nodes $X_1, ... , X_n$, 
\item for each $X_i$, a classical channel $P(X_i^{\textrm{in}}|Pa(X_i)^{\textrm{out}})$ , where 
$Pa(X_i)$ denotes the set of parents of $X_i$ according to $G$, such that $\kappa_{X_1\cdots X_n} = \prod_i P(X_i^{\textrm{in}}|Pa(X_i)^{\textrm{out}})$ is a process operator over $X_1 , ... , X_n$.
\end{enumerate}
\end{definition}
This definition generalizes that of Ref.~\cite{BarrettEtAl_2019_QCMs} to include the case of cyclic graphs, and classical split nodes where the input and output variables have different cardinalities. Ref.~\cite{BarrettEtAl_2019_QCMs} presents detailed discussion of the relationship between (acyclic) CSMs and standard classical causal models \cite{Pearl_Causality, SpirtesEtAL_2000_BookCausationPredictionSearch}. 

In the classical case, causal structure (defined for unitary processes in the quantum case) can be defined for deterministic processes.
\begin{definition} \textnormal{(Causal structure of a deterministic classical process)} \label{Def_detprocess}
Given a deterministic process $\kappa_{X_1 ... X_n}^f$, the \textnormal{causal structure} of the process is the directed graph with vertices $X_1, ..., X_n$ and an arrow $X_i \rightarrow X_j$, whenever $X_j^{\text{in}}$ depends on $X_i^{\text{out}}$ through the function $f$. 
\end{definition}

\begin{definition} \textnormal{(Classical Markov condition --- generalized)} \label{Def_classicalmarkovcondition}
A process $\kappa_{X_1 ... X_n}$ is called \textnormal{Markov} for a directed graph $G$ with classical split-nodes $X_1,\ldots, X_n$ as its vertices iff it admits a factorization of the form $\kappa_{X_1 ... X_n} = \prod_{i=1}^n P(X_i^{\textrm{in}}|Pa(X_i)^{\textrm{out}})$, where $Pa(X_i)$ denotes the set of parents of $X_i$ according to $G$. 
\end{definition}

The following is immediate.
\begin{proposition} \label{Prop_DeterministicProcessMarkov}
Every deterministic classical process is Markov for its causal structure.
\end{proposition}

In the case of general -- i.e., not necessarily deterministic -- classical processes, an account of their relationship to causal structure can be given that again mirrors the quantum case. Let us adopt the provisional approach that causal structure always inheres in deterministic reversible processes (where reversibility here may not be essential, but is assumed to provide a closer analogue to the quantum case in which unitarity is assumed). Then compatibility with a given directed graph can be defined in terms of extension to a reversible deterministic process with latent local noise variables.
\begin{definition} \textnormal{(Reversible extendibility)}
A process $\kappa_{X_1 ... X_n}$ is \textnormal{reversibly extendible} iff there exists a reversible deterministic process $\kappa_{X_1\cdots X_n F \lambda}^f$ with an additional leaf node $F$ and root node $\lambda$, such that $\kappa_{X_1\cdots X_n} = \sum_{F^{\text{in}},\lambda^{\text{out}}} [ \kappa_{X_1\cdots X_n F \lambda}^f P(\lambda^{\text{out}}) ]$  for some $P(\lambda^{\text{out}})$.
\end{definition}

\begin{definition} \textnormal{(Compatibility with a directed graph)} \label{Def_Compatibility}
A process $\kappa_{X_1\cdots X_n}$  is \textnormal{compatible} with a directed graph $G$ with nodes $X_1, ... ,X_n$, iff $\kappa_{X_1\cdots X_n}$  is reversibly extendible to a deterministic process $\kappa_{X_1\cdots X_n F \lambda_1...\lambda_n}^f$, with an additional leaf node $F$, root nodes $\lambda_i$, and a product distribution $\prod_i P(\lambda_i^{\text{out}})$, such that through $f$, $X_i^{\text{in}}$ depends neither on $\lambda_j^{\text{out}}$ for $j \neq i$ nor on $X_j^{\text{out}}$ for $X_j \notin Pa(X_i)$ (with $Pa(X_i)$ referring to $G$).  
\end{definition}

With Prop.~\ref{Prop_DeterministicProcessMarkov}, the following analogue of Thm.~\ref{Thm_CompImpliesMarkov} is straightforward.
\begin{theorem}\label{Thm_classicalcompImpliesmarkov}
If a classical process $\kappa_{X_1\cdots X_n}$ is compatible with a directed graph $G$, then it is also Markov for $G$. 
\end{theorem}
As in the quantum case, we leave open whether the converse to Thm.~\ref{Thm_classicalcompImpliesmarkov} holds. 
\begin{hypothesis} \label{classicalconjecture}
If a process $\kappa_{X_1 ... X_n}$ is Markov for a directed graph $G$, then it is compatible with $G$.
\end{hypothesis}
We remark only that Hypothesis~\ref{classicalconjecture} is not obviously implied by its quantum counterpart, Hypothesis~\ref{quantumconjecture}. First, it is not known whether reversible extendibility implies unitary extendibility for a classical process when seen as a special case of a quantum process. Second, even if this is the case, it is still conceivable that while a classical process that is Markov for a given graph may admit unitary extensions with the required no-influence properties when viewed as a quantum process, no such extension may be equivalent to a deterministic classical process for the given preferred basis. 

We conclude with the following observation.
\begin{theorem} \label{Thm_ClassicalPolytopeResult}
Given a set of classical split nodes $X_1,...,X_n$, the set of reversibly extendible classical processes on $X_1,...,X_n$ coincides with the deterministic polytope.
\end{theorem}
\noindent \textit{Proof:}  See Methods. 

If Hypothesis~\ref{classicalconjecture} holds, then Thm.~\ref{Thm_ClassicalPolytopeResult} implies in particular that the process defined by a CSM must always belong to the deterministic polytope. 
An example of a classical process $\kappa_{X_1\cdots X_n}$ outside of the deterministic polytope is described in Ref.~\cite{Baumeler_EtAl_2016_SpaceOfLOgicallyConsistentClassicalProcesses} (and denoted $\hat{E}_{ex1}$ therein). It is not too hard to show that this process is not Markov for any directed graph, hence cannot be the process defined by a CSM, in keeping with Hypothesis~\ref{classicalconjecture}.

\section*{DISCUSSION}

This work presented an extension of the framework of quantum causal models from Refs.~\cite{AllenEtAl_2016_QCM, BarrettEtAl_2019_QCMs} to include cyclic causal structures.  
We showed that the quantum SWITCH, and a process that violates causal inequalities, found by Ara{\'u}jo and Feix and  described by Baumeler and Wolf, can be seen as the processes defined by cyclic quantum causal models. 
We also gave decompositions of any SWITCH-type process and of the unitary extension of the aforementioned process by Ara{\'u}jo and Feix, enabling diagrammatic representations that make the internal causal structures evident. Applications of these results included proofs that any unitarily extendible bipartite process is causally separable, and that any unitary process is cyclic if and only if it is causally nonseparable. 
 
What technically comes as the natural generalization of the framework of acyclic quantum causal models is conceptually a substantial step -- allowing causal structure to be cyclic. Taking this extended causal model perspective seriously then offers an alternative view of certain processes: a process that is incompatible with definite causal order may now also be seen to have a well-defined cyclic causal structure. This is to say, to admit of a partial order is not an essential property of being causal anymore. While processes that violate a causal inequality were previously referred to as noncausal processes, suggesting they cannot be understood causally, at least some of them then do admit a causal understanding.

Note that as far as acyclic causal structures are concerned there also is the earlier framework of QCMs by Costa and Shrapnel from Ref.~\cite{CostaEtAl_2016_QuantumCausalModeling}, which is related to, in fact strictly contained in that of Refs.~\cite{AllenEtAl_2016_QCM, BarrettEtAl_2019_QCMs}, which the current work extends. The Markov condition of Ref.~\cite{CostaEtAl_2016_QuantumCausalModeling} is a special case of Def.~\ref{Def_MarkovCondition}, restricted to DAGs for which each node's output space factorizes into as many subsystems as the node has children, with each subsystem only influencing the corresponding child. With this idea of a system per arrow, the process operator $\prod_i \rho_{A_i|Pa(A_i)}$ becomes a tensor product. As a consequence -- for essentially the same reason as why Prop.~\ref{Prop_NoBipartite} holds -- the notion of a QCM from Ref.~\cite{CostaEtAl_2016_QuantumCausalModeling} does not admit a nontrivial extension to cyclic directed graphs. The extension of faithful QCMs to cyclic graphs relies on the particular nature of our Markov condition that allows the nontrivial action of pairwise commuting operators $\rho_{A_i|Pa(A_i)}$ to overlap on non-factorizing output spaces.

Although we do not provide the details, we note a further application of the generalized framework: it allows an extended version of the causal discovery algorithm sketched in Ref.~\cite{BarrettEtAl_2019_QCMs} (inspired in turn by the first of its kind in Ref.~\cite{GiarmatziEtAl_2018_CausalDiscoveryAlgorithm}). 
While the version in Ref.~\cite{BarrettEtAl_2019_QCMs}, takes a process operator as input, and outputs DAGs as candidate causal explanations, where possible at all, the extended version can discover and output cyclic causal structures. 
The basic steps of the algorithm in Ref.~\cite{BarrettEtAl_2019_QCMs} largely remain the same, but for instance the algorithm does not halt anymore when encountering a cyclic graph $G_{\sigma}$ that encodes the direct signalling relations between pairs of nodes of the given process $\sigma$. Instead Markovianity for such cyclic $G_{\sigma}$ can still be checked to establish whether $G_{\sigma}$ is a plausible causal explanation.

One of the main questions left open is the validity of our hypothesis that Markovianity implies compatibility for cyclic graphs, which would generalize one of the main results established for the acyclic case in Ref.~\cite{BarrettEtAl_2019_QCMs}. The validity of this hypothesis has consequences, which we spell out as follows.

Ref.~\cite{AraujoEtAl_2017_PurificationPostulateForQMWithIndefiniteCausalOrder}, in motivating the study of unitary extendibility of processes, includes the suggestion that unitary extendibility should be regarded as a necessary condition for a process to be realizable in nature. Here, the meaning of ‘realizable’ is a little vague, but might be taken, for example, to include exotic scenarios involving gravity as well as the time-delocalized sense discussed above in which some processes have been realized in the laboratory. (It does not include realization via postselection, since it is known that all processes can be realized under a suitable postselection \cite{OreshkovEtAl_2016_OperationalQuantumTheoryWithoutTime, SilvaEtAl_2017_ConnectingIndefiniteWithMultiTime, AraujoEtAl_2017_QuantumComputationWithIndefiniteCausalStructure, MilzEtAl_2018_EntanglementNonMarkovianityAndCausalNonSeparability}.) The suggestion would hold if all processes, once sufficient systems are included, are unitary at the most fundamental level.

Alternatively, under the assumption that the process operator framework provides the most general description of the possible correlation between quantum systems, in non-postselected scenarios, one may speculate that a necessary condition for a process to be realizable in nature is that it can arise from a QCM. Here, ‘arise’ means that there is a QCM with process $\sigma’$ such that $\sigma$ can be obtained from $\sigma’$ by inserting channels at some of the nodes of $\sigma’$ and marginalizing over them. The idea is that any correlations described by such a process admit a causal explanation, albeit one that may involve cycles. On the other hand, any process that cannot arise from a QCM in this manner describes correlations that are not amenable to an understanding in causal terms.

The connection with unitary extendibility is that any process that is unitarily extendible has the property that it can arise from a QCM. Furthermore, if Hypothesis~\ref{quantumconjecture} holds, then any process that is not unitarily extendible cannot arise from a QCM. Hence if Hypothesis~\ref{quantumconjecture} holds, the speculation above coincides with the suggestion of Ref.~\cite{AraujoEtAl_2017_PurificationPostulateForQMWithIndefiniteCausalOrder}.

If Hypothesis~\ref{quantumconjecture} fails, there is a peculiar class of cyclic quantum causal models, in which the process is Markov for the graph but not compatible with the graph. 
There then are two logically conceivable options: one may insist on the notion of compatibility as the essential concept for giving causal explanations, turning the Markov condition into a necessary but insufficient condition;  alternatively, one could insist on the Markov condition as the essential concept for giving causal explanations, turning the current notion of compatibility into a sufficient but not necessary condition.  
We leave open the question whether any meaning can be given to the arrows of the graph in this case, given that there is no suitable unitary extension to define causal relations, and whether such processes might be realizable or not. 

Beyond establishing the hypothesis, future work might study the extent to which other core results of the framework of quantum causal models in the acyclic case, such as the d-separation theorem \cite{BarrettEtAl_2019_QCMs}, can be generalized in an appropriate way to the cyclic case, as has been done for the classical framework (see, e.g., Ref.~\cite{ForreEtAl_2019_CausalCalculusWithCycles}). 

Finally, one of the most promising avenues for future work is the general idea behind the above causal decompositions of our example processes together with Thm.~\ref{Thm_ResultCausalSep}\hspace*{0.07cm}: 
to derive further causal decompositions of unitary transformations $U$, as started in Ref.~\cite{LorenzEtAl_2020_CausalAndCompStructure}, and then study the interplay between the discovered algebraic structure and the condition that $U$ defines a valid unitary process when identifying in- and output spaces of $U$ as the out- and input spaces of quantum nodes. We expect this  mathematical tool to lead to insights into which unitarily extendible processes are causally nonseparable and how the cyclicity is distributed, mathematically speaking, across the process -- with possible hints for the process' physical realizability.

\section*{METHODS}

\subsection*{Characterisation of process operators} 

In order state necessary and sufficient conditions for an operator to be a valid process operator, the following will be useful. 
Let $\{\eta_X^l\}_{l=0}^{d_X^2-1}$ denote a Hilbert-Schmidt (HS) basis for $\mathcal{L}(\mathcal{H}_{X})$, i.e., a set of operators such that they are orthonormal with respect to the HS inner product and, in addition, traceless for all $l=1,...,d_X^2-1$, while $\eta_X^0=(1/d_X) \mathds{1}_{X}$.
Any $\sigma \in \mathcal{L}( \mathcal{H}_{A^{\text{in}}} \otimes  \mathcal{H}_{A^{\text{out}}} \otimes \mathcal{H}_{B^{\text{in}}} \otimes  \mathcal{H}_{B^{\text{out}}} )$ can be expanded in a HS basis as $\sigma = \sum_{l_1,l_2,l_3,l_4} \ \alpha_{l_1l_2l_3l_4} \ 
\eta^{l_1}_{A^{\text{in}}} \otimes  \eta^{l_2}_{A^{\text{out}}} \otimes \eta^{l_3}_{B^{\text{in}}} \otimes  \eta^{l_4}_{B^{\text{out}}}$. 
A term of type $A^{\text{in}}$ in the expansion is a summand with non-trivial action only on $A^{\text{in}}$, i.e. $l_1\neq 0$ and $l_2=l_3=l_4=0$. Similarly for types $A^{\text{in}}B^{\text{out}}$ etc.

It was shown in Ref.~\cite{OreshkovEtAl_2012_QuantumCorrelationsWithoutCausalOrder} that $\sigma$ being a bipartite process operator is equivalent to $\sigma \geq 0$, $\Trace [\sigma] = d_{A^{\text{out}}} d_{B^{\text{out}}} $ and that in a HS basis expansion, in addition to a term, which is proportional to the identity operator on all four spaces, only the coefficients of terms of the types 
$A^{\text{in}}$, $B^{\text{in}}$, $A^{\text{in}}B^{\text{in}}$, 
$A^{\text{in}}B^{\text{out}}$, $A^{\text{out}}B^{\text{in}}$, 
$A^{\text{in}}A^{\text{out}}B^{\text{in}}$ and $A^{\text{in}}B^{\text{in}}B^{\text{out}}$, 
may be non-vanishing. 
These conditions were generalized to $n$ numbers of nodes in Ref.~\cite{OreshkovEtAl_2016_CausallySeparableProcesses} and can easily be stated as (1) $\sigma \geq 0$, (2) $\Trace [\sigma] =\prod_{i=1}^n d_{A_i^{\text{out}}}$ and (3) that in a HS basis expansion the only non-vanishing terms, apart from an overall identity operator, are of a type such that there must be at least one node, say $A_i$, on whose out-space, $A_i^{\text{out}}$, the action is trivial, but on whose in-space, $A_i^{\text{in}}$, the action is non-trivial.
Equivalent conditions were presented in \cite{AraujoEtAl_2015_WitnessingCausalNonSeparability} where the projector onto the linear subspace of process operators was defined explicitly, giving a basis-independent characterization.

\subsection*{Proof of Prop.~\ref{Prop_NoBipartite}}  

Suppose a bipartite cyclic QCM is given by the (unique) cyclic graph $G$ with two nodes $A$ and $B$ from Fig.~2a and a process $\sigma_{AB} = \rho_{A|B} \ \rho_{B|A}$, Markov for $G$. It follows that $\sigma_{AB} = \rho_{B|A} \otimes \rho_{A|B}$, as both factors act on distinct Hilbert spaces. Now suppose that this is a faithful QCM, i.e., both channels $\rho_{A|B}$ and $\rho_{B|A}$ are signalling channels. One way to see that this contradicts the assumption that $\sigma_{AB}$ is a valid process is by analyzing the non-vanishing types of terms in an expansion of $\sigma_{AB}$ relative to a Hilbert-Schmidt product basis \color{black} (see above). \color{black} 
If signalling from $B^{\text{out}}$ to  $A^{\text{in}}$ is possible in $\rho_{A|B}$, then an expansion of just  $\rho_{A|B}$ has to contain a non-vanishing term of type $A^{\text{in}}B^{\text{out}}$. Similarly, if signalling from $A^{\text{out}}$ to  $B^{\text{in}}$ is possible in $\rho_{B|A}$, then an expansion of $\rho_{B|A}$ has to contain a non-vanishing term of type $B^{\text{in}}A^{\text{out}}$. Consequently, $\sigma_{AB}$ has to contain a non-vanishing term of type $A^{\text{in}}B^{\text{out}}B^{\text{in}}A^{\text{out}}$, which is forbidden for a process operator \cite{OreshkovEtAl_2012_QuantumCorrelationsWithoutCausalOrder}.

\subsection*{Product of commuting operators not necessarily a process operator}

As established by Prop.~\ref{Prop_NoBipartite}, not all cyclic graphs support a faithful cyclic QCM. 
Here we show that, \color{black} given a cyclic graph $G$ that does support a faithful cyclic QCM, it is not true that any product of commuting operators $\prod_i \rho_{A_i|Pa(A_i)}$, with parental sets as in $G$, constitutes a process operator.  
Consider for instance the graph $G$ in Fig.~2b 
(and see \color{black} the discussion below Def.~\ref{Def_CausalSep} for an example of \color{black} a faithful cyclic QCM over $G$). 
Letting the three nodes $A$, $B$ and $C$ be \color{black} classical split nodes, \color{black} with classical bits $A^{\text{in}}$, $A^{\text{out}}$, $B^{\text{in}}$, $B^{\text{out}}$, $C^{\text{in}}$ and $C^{\text{out}}$, define classical channels as in Eqs.~\eqref{Eq_CounterExampleDistr1}-\eqref{Eq_CounterExampleDistr2}. 
It is easy to see that the signalling relations through the channels $P(A^{\text{in}}|B^{\text{out}},C^{\text{out}})$ and $P(B^{\text{in}}|A^{\text{out}},C^{\text{out}})$ are indeed as in Fig.~2b. 
At the same time, for any choice of probability distribution $P(C^{\text{in}})$, the product 
$P(A^{\text{in}}|B^{\text{out}}, C^{\text{out}}) P(B^{\text{in}}|A^{\text{out}}, C^{\text{out}}) P(C^{\text{in}})$ cannot be a classical process: consider an intervention at $C$ which fixes $C^{\text{out}}$ to be 0, then 
$P(A^{\text{in}}|B^{\text{out}}, 0) P(B^{\text{in}}|A^{\text{out}}, 0)$ is still a product of two signalling classical channels, which (seeing them as special cases of quantum channels) was already established in the proof of Prop.~\ref{Prop_NoBipartite} to be in contradiction with being a process. This establishes the claim. 

\begin{widetext}
	\begin{eqnarray}
		P(A^{\text{in}}|B^{\text{out}},C^{\text{out}}) &:=& 
		\begin{cases} 
			P(0|0,0) \ = \ 0.4, \hspace*{0.5cm} 
			P(0|0,1) \ = \ 0.3, \hspace*{0.5cm} 
			P(0|1,0) \ = \ 0.8, \hspace*{0.5cm} 
			P(0|1,1) \ = \ 0.3, \\
			P(1|0,0) \ = \ 0.6, \hspace*{0.5cm} 
			P(1|0,1) \ = \ 0.7, \hspace*{0.5cm} 
			P(1|1,0) \ = \ 0.2, \hspace*{0.5cm} 
			P(1|1,1) \ = \ 0.7. 
		\end{cases}
		\label{Eq_CounterExampleDistr1} \\
		P(B^{\text{in}}|A^{\text{out}},C^{\text{out}}) &:=& 
		\begin{cases} 
			P(0|0,0) \ = \ 0.5, \hspace*{0.5cm} 
			P(0|0,1) \ = \ 0.3, \hspace*{0.5cm} 
			P(0|1,0) \ = \ 0.25, \hspace*{0.5cm} 
			P(0|1,1) \ = \ 0.1, \\
			P(1|0,0) \ = \ 0.5, \hspace*{0.5cm} 
			P(1|0,1) \ = \ 0.7, \hspace*{0.5cm} 
			P(1|1,0) \ = \ 0.75, \hspace*{0.5cm} 
			P(1|1,1) \ = \ 0.9. 
		\end{cases}
		\label{Eq_CounterExampleDistr2}
	\end{eqnarray}
\end{widetext}

\subsection*{Proof of Thm.~\ref{Thm_ResultCausalSep}}

Suppose the bipartite quantum process operator $\sigma_{AB}$ is unitarily extendible. Consider an arbitrary unitary extension of it, $\sigma_{ABFP}=\rho^{\mathcal{U}}_{ABF|ABP}$. From Eq.~\eqref{subsystemdecomposition} it follows that the reduced process obtained by tracing out $F^{\text{in}}$ has the form 
\begin{equation}
\sigma_{ABP} = \Trace_{F^{\text{in}}} [\rho^{\mathcal{U}}_{ABF|ABP}] = \sum_{i\in I} \rho_{A|BP_i^L} \otimes \rho_{B|P_i^RA} \ ,  
\end{equation}
for the decomposition $\mathcal{H}_{P^{\text{out}}} = \bigoplus_{i\in I}  \mathcal{H}_{P_i^L} \otimes \mathcal{H}_{P_i^R}$, identified by $S$, where 
$\rho_{A|BP_i^L}= \Trace_{F^L_i} [ \rho^{V_i}_{AF^L_i|BP_i^L}]$ and $\rho_{B|P_i^RA}= \Trace_{F^R_i} [\rho^{W_i}_{F^R_iB|P_i^RA}]$ and, where $\rho_{A|BP_i^L} \otimes \rho_{B|P_i^RA}$ is taken as an operator on the whole space, acting as zero map on all but the $i$th subspace.   
Note that from $\sigma_{ABP}$ being a process operator it follows that feeding in any $\tau_P \in \mathcal{L}(\mathcal{H}^*_{P^{\text{out}}})$ gives a quantum process operator on the nodes $A$ and $B$. 
Let $i\in I$ be some fixed index and suppose through the channel  $\rho_{A|BP_i^L}$ system $B^{\text{out}}$ can signal to $A^{\text{in}}$ and similarly, through the channel $\rho_{B|AP_i^R}$ system $A^{\text{out}}$ can signal to $B^{\text{in}}$. 
Then there exists an appropriate state $\tau_P$, which has only support on the $i$th subspace, and which is of a product form $\gamma_{P_i^L} \otimes \phi_{P_i^R}$, such that in 
\begin{equation}
	\Trace_{(P_i^L)^*}[ \rho_{A|BP_i^L} \ \gamma_{P_i^L} ]  \ \ \otimes  \ \ \Trace_{(P_i^R)^*}[  \rho_{B|AP_i^R} \ \phi_{P_i^R}] \ , \label{Eq_MarginalIniSubspace}
\end{equation}
both, the marginal channel on the left is signalling from $B^{\text{out}}$ to $A^{\text{in}}$ and the one on the right from $A^{\text{out}}$ to $B^{\text{in}}$. 
Since the expression in Eq.~\eqref{Eq_MarginalIniSubspace} has to give a process operator over $A$ and $B$, this yields a contradiction due to Prop.~\ref{Prop_NoBipartite}. Hence, for each $i$ at most one of the channels $\rho_{A|BP_i^L}$ and $\rho_{B|AP_i^R}$ allow signalling from $B^{\text{out}}$ to $A^{\text{in}}$ or from $A^{\text{out}}$ to $B^{\text{in}}$, respectively. 
By assumption there exists an appropriate $\tau_P \in \mathcal{L}(\mathcal{H}^*_{P^{\text{out}}})$ such that 
\begin{equation}
\sigma_{AB} =  \sum_i \Trace_{(P^{\text{out}})^*} \Big[ (\rho_{A|BP_i^L} \otimes \rho_{B|P_i^RA} ) \  \tau_P \Big] \ . \label{Eq_CS} 
\end{equation}
By the above analysis, it also follows that each summand in Eq.~\eqref{Eq_CS} has to be a process operator up to normalization. 
Since they sum up to a process operator, the inverses of the normalization constants have to form a probability distribution and one can therefore write $\sigma_{AB} =  \sum_i p_i \ \sigma_{AB}^{(i)}$, where each $\sigma_{AB}^{(i)}$ is a process operator with at most $A$ signalling to $B$ or vice versa. This is the form of a bipartite causally separable process operator. 

Note further that if $\rho_{A|BP_i^L}$ is non-signalling from $B^{\text{out}}$ to $A^{\text{in}}$, then in $V_i$ there is no influence from $B^{\text{out}}$ to $A^{\text{in}}$, and similarly, if $\rho_{B|P_i^RA}$ is non-signalling from $A^{\text{out}}$ to $B^{\text{in}}$, then in $W_i$ there is no influence from $A^{\text{out}}$ to $B^{\text{in}}$. 
Therefore, the above constraints mean that each term $V_i\otimes W_i$ in Eq.~\eqref{subsystemdecomposition} corresponds to a process over nodes including $A$ and $B$ that allows signalling in at most one direction between $A$ and $B$. 
The latter always admits an implementation as a unitary circuit fragment 
with nodes $A$ and $B$ in a fixed order \cite{ChirbiellaEtAl_2009_QuantumNetworkFramework}. Since the full unitary $U$ of the unitary extension is a direct sum of such fixed-order unitary processes taking place in the different orthogonal subspaces, and every operation at the nodes $A$ and $B$ can be dilated to a unitary, the full unitary process $\sigma_{ABFP}=\rho^U_{ABF|ABP}$ can be realized by coherently conditioning which of the corresponding fixed-order unitary circuits takes place on the logical value of some control $n$-level quantum system, where $n$ is the number of different subspaces. 
Note that since the systems involved in the fixed-order circuits may have different dimensions, this implementation in practice may require bringing in different systems depending on the control variable $i$, but this can always be seen as part of a process on a larger system of a fixed dimension. 
Moreover, the fixed-order processes in the different orthogonal subspaces can be grouped into two sets: one in which $A$ is before $B$ and another one in which $B$ is before $A$. 
This allows embedding the process into another one where one of two possible circuits (in which $A$ and $B$ occur in different orders) is applied in a coherently controlled fashion based on the logical value of a control qubit, similarly to the quantum SWITCH. 
This yields another possible unitary extension $\sigma_{AB\widetilde{F}\widetilde{P}}$ of the original bipartite process, where $\widetilde{F}^{\text{in}}$ and $\widetilde{P}^{\text{out}}$ would contain $F^{\text{in}}$ and $P^{\text{out}}$, respectively, as subspaces. 
The originally assumed unitary extension $\sigma_{ABFP}$ can then be seen to take place effectively as part of $\sigma_{AB\widetilde{F}\widetilde{P}}$.

\subsection*{Proof of Thm.~\ref{Thm_CyclicityAndNonSep}} 

The below proof of Thm.~\ref{Thm_CyclicityAndNonSep} will use the following two concepts. 
First, generalizing the notion of a process being unitary, a process is called isometric if its induced channel from the output systems of all nodes to the input systems of all nodes arises from an isometry. 
Second, a quantum comb, as defined in Ref.~\cite{ChirbiellaEtAl_2009_QuantumNetworkFramework} (provided first input and last output system are trivial), is a special kind of quantum process: a process $ \sigma_{A_{1} \ldots A_{n}}$ over $n$ quantum nodes for the given total order of its nodes $A_{1}, \ldots, A_{n}$ is a quantum comb (an $(n+1)$-comb) iff  
\begin{eqnarray}
	\ \forall l = 1, \ldots, n-1 \hspace*{0.3cm} 
	\textrm{Tr}_{A_{l+1} \ldots  A_{n}} [ \sigma_{A_{1} \ldots A_{n}} ] =  \hspace*{1.7cm} \nonumber \\
	\hspace*{0.45cm}  \frac{1}{d_{A_{l}^{\textrm{out}}}} \textrm{Tr}_{(A_{l}^{\textrm{out}})^*} \Big[ \textrm{Tr}_{ A_{l+1} \ldots  A_{n}} [ \sigma_{A_{1} \ldots A_{n}} ] \Big] 
		\otimes \mathds{1}_{(A_{l}^{\textrm{out}})^*}  \hspace*{0.35cm} \label{Eq_DefQuantumComb} \\ 
		\wedge \ \sigma_{A_{1} \ldots A_{n}} =  \frac{1}{d_{A_{n}^{\textrm{out}}}} \textrm{Tr}_{(A_{n}^{\textrm{out}})^*} [ \sigma_{A_{1} \ldots A_{n}} ] \otimes \mathds{1}_{(A_{n}^{\textrm{out}})^*} . \hspace*{0.6cm} \label{Eq_DefQuantumComb_2}
\end{eqnarray}

\noindent \textit{Proof of Thm.~\ref{Thm_CyclicityAndNonSep}:} 
Let $\sigma_{A_1\ldots A_n}$ be a unitary process. 
The following will establish, what is equivalent to Thm.~\ref{Thm_CyclicityAndNonSep}, namely that acyclicity of its causal structure is equivalent to $\sigma_{A_1\ldots A_n}$ being causally separable. 

First, suppose $\sigma_{A_1\ldots A_n}$ has an acyclic causal structure. 
There then exists a total order of the quantum nodes $A_1, \ldots, A_n$ (appropriately relabeled) such that 
$A_j \nrightarrow A_i$ $\forall j \geq i$ (see Def.~\ref{Def_UnitaryProcess}).  
This implies that the conditions in Eqs.~\eqref{Eq_DefQuantumComb}-\eqref{Eq_DefQuantumComb_2} are satisfied (note that $d_{A_{n}^{\textrm{out}}} = 1 = d_{A_{1}^{\textrm{in}}}$). 
Hence, $\sigma_{A_1 \ldots A_n}$ is a quantum comb \cite{ChirbiellaEtAl_2009_QuantumNetworkFramework}. Such a process is a special case of a causally separable process since in a quantum comb there can be no signalling from $\{A_{j+1}, \cdots, A_{n} \} $ to $ \{ A_{1}, \cdots, A_{j}\}$ for any $j = 1, \cdots, n-1$, and this remains true under extending the process with arbitrary shared input ancillary states. 

For the converse direction, suppose the unitary process $\sigma_{A_1\ldots A_n}$ is causally separable. 
In order to show that it then has an acyclic causal structure we will prove that it is a quantum comb. In fact we will prove the following more general statement concerning isometric processes, which gives the claim as a special case. 

\begin{lemma} \label{Lem_CausSepIsometries}
	Every causally separable isometric process is a quantum comb.
\end{lemma}

\noindent \textit{Proof of Lem.~\ref{Lem_CausSepIsometries}:} 
The main idea of the following proof is the observation that the process operator of an isometric process is proportional to a rank-1 projector and hence cannot be written as a nontrivial convex mixture of different positive semi-definite operators. 
The proof proceeds by induction.

An isometric process over one single node is a 2-comb. 
Assume that all causally separable isometric processes on $n$ nodes are quantum combs.  
Let $\sigma_{A_1\ldots A_{n+1}}$ be an isometric process over $n+1$ nodes, which is causally separable. 
Let us extend it by adding auxiliary input systems for all $n+1$ nodes with the following pure state shared among them: 
\begin{gather}
|\Psi\rangle \ = \ \bigotimes_{i=1,j=2, i<j}^{i=n, j=n+1} \   |\phi^+\rangle_{ij} \ , \label{Eq_ancillarystate}
\end{gather}
where each $|\phi^+\rangle_{ij} = \frac{1}{\sqrt{n!}} \sum_{l=1}^{n!} |l\rangle |l\rangle$ 
is a maximally entangled state, shared between node $A_{i}$ and node $A_{j}$. Thus, $|\Psi\rangle$ is a tensor product of $\frac{1}{2}n(n+1)$ maximally entangled bipartite states, such that every pair of nodes indexed by $(i,j)$ shares one such state of Schmidt rank $n!$. 
Using the notation of Def.~\ref{Def_CausalSep}, $\widetilde{\sigma} :=  \sigma \otimes |\Psi\rangle\langle\Psi|$ is an extended process over the extended nodes $\widetilde{A}_1, \ldots, \widetilde{A}_{n+1}$, with $|\Psi\rangle \in \bigotimes_{i=1}^{n+1} \mathcal{H}_{(A'_i)^{\text{in}}}$, where each $\mathcal{H}_{(A'_i)^{\text{in}}}$ is an $n$-fold tensor product of $(n!)$-dimensional systems. 

By assumption, $\widetilde{\sigma}$ is causally separable, too, while it also is proportional to a rank-1 projector. 
In the decomposition as in Eq.~\eqref{Eq_CSdecomposition}, implied by causal separability, there therefore is only one summand. Hence, there exists one node, let this be $\widetilde{A}_{1}$ (for an appropriate relabeling), such that  $\widetilde{A}_2, \ldots, \widetilde{A}_{n+1}$ cannot signal to $\widetilde{A}_{1}$ and for all CP maps $\tau_{\widetilde{A}_{1}}$ at that node the conditional process 
$\widetilde{\sigma}|_{\tau_{\widetilde{A}_1}}$ is causally separable. 
Now consider a CP map such that $\tau_{\widetilde{A}_{1}} =  |\tau\rangle \langle \tau|_{\widetilde{A}_{1}}$ itself is a rank-1 projector. The process operator $\widetilde{\sigma}|_{\tau_{\widetilde{A}_1}}$ then still is proportional to a rank-1 projector and, hence, representing an isometric process on the remaining $n$ nodes $\widetilde{A}_2, \ldots, \widetilde{A}_{n+1}$. 
As argued above it also is causally separable. 
By assumption then such an isometric, causally separable process $\widetilde{\sigma}|_{\tau_{\widetilde{A}_1}}$ on $n$ nodes is a quantum comb.  

Notice first that if there is no signalling to  $\widetilde{A}_{1}$ from all other nodes in the extended process $\widetilde{\sigma}$, then there is no signalling to $A_{1}$ from all other nodes in the original process $\sigma$. 
Consider $\tau_{\widetilde{A}_{1}}  = |\tau\rangle \langle \tau|_{{A}_{1}} \otimes |\phi\rangle\langle \phi|$, 
where $ |\phi\rangle\langle \phi|$ is some fixed projector on the ancillary input system $(A'_1)^{\text{in}}$ and 
$\tau_{A_{1}} = |\tau\rangle \langle \tau|_{{A}_{1}} $ has rank-1. 
Since projecting the ancillary systems via $|\phi\rangle\langle \phi|$ leaves the ancillary systems on the remaining nodes in some pure state $|\Phi\rangle\langle \Phi|$, the conditional process on the remaining nodes has the form 
$\sigma|_{\tau_{A_{1}}} \otimes |\Phi\rangle\langle \Phi|$. Since the latter is a quantum comb for every $|\tau\rangle \langle \tau|_{{A}_{1}}$, so must be $\sigma|_{\tau_{A_{1}}}$. 

There are $n!$ different possible total orders of the nodes, given by $A_{\pi(2)}, \ldots, A_{\pi(n+1)}$ for $\pi$ being one of the $n!$ different permutations of $2,...,n+1$. 
We will now show (by proof of contradiction) that there exists a reordering $A_{\pi(2)}, \ldots, A_{\pi(n+1)}$ with which the quantum comb 
$\sigma|_{\tau_{A_1}}$ is compatible for any choice of $|\tau\rangle \langle \tau|_{{A}_{1}}$. 
Suppose there does not exist one such appropriate total order. 
Then for every permutation $\pi$, there exists $\tau^{\pi}_{A_{1}} := |\tau^{\pi}\rangle \langle \tau^{\pi}|_{{A}_{1}}$, such that the corresponding quantum comb $\sigma|_{\tau^{\pi}_{A_{1}}}$ is incompatible with the total order of the remaining nodes defined by $\pi$.  
Let $\mathcal{C}_{l}^{\pi} (\sigma) = 0$ for $l=1,...,n$ be the linear constraint corresponding to the $l$th condition in Eqs.~\eqref{Eq_DefQuantumComb}-\eqref{Eq_DefQuantumComb_2} for a process operator $\sigma$ over $n$ nodes to be a valid quantum comb for the total order $\pi$. 

Consider a process operator 
$\bar{\sigma} := \sum_{\pi=1}^{n!} q_{\pi} \ \sigma|_{\tau^{\pi}_{A_{1}}}$, where $q_{\pi} \geq 0$, $ \forall \pi$, and $\sum_{\pi} q_{\pi} =1$ (letting $\pi$, both, be a permutation as well as an index enumerating those permutations). 
By construction, for every $\pi$ at least one of the conditions in $\{ \mathcal{C}_{l}^{\pi} ( \sigma|_{\tau^{\pi}_{A_{1}}} ) = 0 \}_{l=1}^n$ fails. 
Therefore, one can then choose the weights $q_{\pi}$ such that for every $\pi$ the process operator $\bar{\sigma}$ violates at least one of these constraints $\{ \mathcal{C}_{l}^{\pi} ( \bar{\sigma} ) = 0 \}_{l=1}^n$, establishing that $\bar{\sigma}$ is not a quantum comb for any possible order of the $n$ nodes. 
More precisely, the condition that $\bar{\sigma}$ respects the constraints 
$\{ \mathcal{C}_{l}^{\pi} ( \bar{\sigma} ) = 0 \}_{l=1}^n$, for a given $\pi$ can be written as 
$\sum_{\alpha=1}^{n!} q_{\alpha} \ \mathcal{C}_{l}^{\pi} ( \sigma|_{\tau^{\alpha}_{A_{1}}} )= 0$ for $l=1,\ldots, n$, which implies that 
$(q_1, \ldots, q_{n!})$, viewed as a point in an $(n!)$-dimensional Euclidean space, must belong to a specific hyperplane in that space. Our assumption that at least one of $\mathcal{C}_{l}^{\pi} ( \sigma|_{\tau^{\pi}_{A_{1}}} )$ must be nonzero, makes it a proper hyperplane. Then, in order for $\bar{\sigma}$ to be compatible with the quantum-comb conditions for at least one $\pi$, the point $(q_1, \ldots, q_{n!})$ must belong to the union of the hyperplanes corresponding to the different values of $\pi$. Since this is a finite set of hyperplanes, it is possible to find (a continuum of) points in the positive orthant that are outside of this union. Since rescaling $(q_1, \ldots, q_{n!})$ by a constant factor, which amounts to rescaling $\bar{\sigma}$ by a constant factor, does not change the fact of whether any of the above constraints is violated or not, there exists a $(q_1, \ldots, q_{n!})$ with the required properties, such that $\bar{\sigma}$ is not a quantum comb for any total order $\pi$. 

We will now use this fact to construct the contradiction with the assumption that there is no single order $\pi$ with which all isometric quantum combs $\sigma|_{\tau_{A_{1}}}$ are compatible. 
To this end, we will first show that, starting from our extended process $\widetilde{\sigma} =  \sigma \otimes |\Psi\rangle\langle\Psi|$, for any $j\in \{2,...,n+1\}$ it is possible to apply a suitable CP map $|\tau\rangle\langle \tau|_{\widetilde{A}_{1}}$ such that this yields a conditional process of the form 
$\widetilde{\sigma}|_{ \tau_{\widetilde{A}_{1}}} =  |  \bar{\sigma}_j\rangle\langle  \bar{\sigma}_j |  \otimes |\Phi\rangle\langle \Phi|_{rest^{'}_{\textrm{in}} }$, where 
$|\bar{\sigma}_j \rangle= \sum_{\pi=1}^{n!} \sqrt{q_{\pi} } \ |\pi\rangle_{a_{j}} | \sigma_{\tau^{\pi}_{A_{1}}} \rangle$ with, recalling Eq.~\eqref{Eq_ancillarystate},  
$\mathcal{H}_{a_{j}}$ the factor of $\mathcal{H}_{(A'_j)^{\text{in}}}$ sharing the state $|\phi^+\rangle_{1j}$ with $\mathcal{H}_{a_{1}}$ of the node $A_1$ and 
$| \sigma_{\tau^{\pi}_{A_{1}}} \rangle \langle \sigma_{\tau^{\pi}_{A_{1}}} | := \sigma|_ {\tau^{\pi}_{A_{1}}}$, the conditional process on the remaining $n$ of the original $n+1$ nodes, and where $|\Phi\rangle\langle \Phi|_{rest^{'}_{\textrm{in}}}$ is some pure state on the remaining auxiliary input systems (i.e. $|\Phi\rangle_{rest^{'}_{\textrm{in}}}$ is in $\bigotimes_{i\neq 1} \mathcal{H}_{(A'_i)^{\text{in}}}$ excluding the subfactor $\mathcal{H}_{a_{j}}$). 

To see this, let $j \neq 1$. 
If we apply a CP map of the form $|\tau\rangle\langle \tau|_{\widetilde{A}_{1}}$ = $ |\chi\rangle \langle\chi|_{a_{1}A_{1}}   \otimes |\phi\rangle \langle \phi|_{rest_{\widetilde{A}_{1}}}$, where 
$|\chi\rangle = \sum_{\pi=1}^{n!}  \sqrt{\epsilon_{\pi}} \ | \pi \rangle_{a_{1} } | \tau^{\pi}\rangle_{A_{1}}$, and 
$|\phi\rangle \langle \phi|_{rest_{\widetilde{A}_{1}}}$ is some projector on the remaining ancillary input systems in 
$(A'_1)^{\text{in}}$, then we will obtain a conditional process of the form 
$\widetilde{\sigma}|_{\tau_{\widetilde{A}_{1}}} =  | {\sigma}_j\rangle\langle   {\sigma}_j |  \otimes |\Phi\rangle\langle \Phi|_{rest^{'}_{\textrm{in}} }$, with 
$ |  {\sigma}_j \rangle= \frac{1}{ \sqrt{\sum_{\pi=1}^{n!} \epsilon_{\pi}\gamma_{\pi}  } } \sum_{\pi=1}^{n!}    \sqrt{\epsilon_{\pi} \gamma_{\pi }} |\pi\rangle_{a_{j}} | \sigma_{\tau^{\pi}_{A_{1}}} \rangle$, where 
$\gamma_{\pi} := {\textrm{Tr} [(|\tau^{ \pi} \rangle\langle  \tau^{\pi}|_{A_{1}}) \sigma ]}$. 
Therefore, by choosing $\epsilon_{\pi} = q_{\pi}/ (c \gamma_{ \pi})$, for some large enough constant $c$ to ensure that $|\tau\rangle\langle \tau|_{\widetilde{A}_{1}}$ is appropriately normalised to represent a CP map, we can make $ | {\sigma}_j\rangle =  |  \bar{\sigma}_j\rangle$ as desired. (Note that $\forall \pi$, $\gamma_{ \pi} \neq 0$ since $\textrm{Tr}_{A_1} [(|\tau^{ \pi} \rangle\langle  \tau^{\pi}|_{A_{1}}) \sigma ]$ is proportional to a process operator on the remaining $n$ nodes, the trace over which gives $\prod_{i=2}^{n+1} d_{A_i^{\text{out}}}$.)

By our main assumption, the $n$-node process $\widetilde{\sigma}|_{\tau_{\widetilde{A}_{1}}} =  | \bar{\sigma}_j\rangle\langle \bar{\sigma}_j |  \otimes |\Phi\rangle\langle \Phi|_{rest^{'}_{\textrm{in}} }$ must be a quantum comb, and since $|\Phi\rangle\langle \Phi|_{rest^{'}_{\textrm{in}} }$ is just a state on some input systems, $|  \bar{\sigma}_j\rangle\langle   \bar{\sigma}_j | $ must also be a quantum comb (on the nodes $A_i\neq A_{1}$, $i\neq j$, and the node $A_j$ extended via the ancillary input system $a_j$). 
But tracing out the system $a_j$ from the latter quantum comb must also yield a quantum comb on the nodes $A_i\neq A_{1}$, which can easily be seen from the quantum-comb conditions.  
However, by construction, $\textrm{Tr}_{a_j} |\bar{\sigma}_j\rangle\langle \bar{\sigma}_j |= \bar{\sigma}$, where $\bar{\sigma}$ is not supposed to be a quantum comb, which is a contradiction. 

Therefore, there must exist a total order $\bar{\pi}$, such that $\sigma|_{ \tau_{A_{1} }}$ is a quantum comb compatible with $\bar{\pi}$ for every rank-1 $\tau_{A_{1}}$. By the convexity of the set of $n$-node operators that are quantum combs compatible with $\bar{\pi}$, this automatically extends to all CP maps  $\tau_{A_{1}}$. 

So far we have shown that the process $\sigma$ is such that there is a node $A_{1}$ to which the rest of the nodes cannot signal, and the remaining nodes can be put in a total order $A_{2}, \ldots , A_{n+1}$, such that for every CP map $\tau_{A_{1}}$, the conditional process $\sigma|_{ \tau_{A_{1} }}$ is a quantum comb compatible with that order. 
Now observe that this implies that the full process  $\sigma$ is a quantum comb compatible with the total order $A_{1}, A_{2}, \ldots , A_{n+1}$. 
Since for all possible CP maps $\tau_{A_{1}}$ it holds that $\mathcal{C}_{l}(\sigma|_{ \tau_{A_{1}}})= 0$ for $l=2,...,n+1$, it follows from the linearity of these constraints, that the corresponding quantum comb conditions hold for $\sigma$, i.e. $\mathcal{C}_{l}(\sigma)= 0$ for $l=2,...,n+1$. 
Finally, that $\mathcal{C}_{1}(\sigma)= 0$ holds follows from just $\sigma$ being a process, since it is equivalent to that if in $\sigma$ we trace out all of the nodes $A_{2}, \ldots , A_{n+1}$, we should be left with, up to normalization, a valid single-node process on $A_{1}$ \cite{OreshkovEtAl_2012_QuantumCorrelationsWithoutCausalOrder}.
Therefore, the isometric process $\sigma$ on $n+1$ nodes is a quantum comb, too, which completes the proof of Lem.~\ref{Lem_CausSepIsometries} and thereby also that of Thm.~\ref{Thm_CyclicityAndNonSep}.

\subsection*{Proof of Thm.~\ref{Thm_ClassicalPolytopeResult}} 

First, suppose $\kappa_{X_1...X_n}$ is a reversibly extendible process, that is, there exists a reversible deterministic process $\kappa^g_{X_1...X_n \lambda F}$ for some bijection 
$g : X_1^{\text{out}} \times ... \times X_n^{\text{out}} \times \lambda^{\text{out}} \rightarrow  X_1^{\text{in}} \times ... \times X_n^{\text{in}}  \times F^{\text{in}}$, such that 
\begin{equation}
	\kappa_{X_1...X_n} = \sum_{\lambda^{\text{out}}, F^{\text{in}}} \kappa^g_{X_1...X_n \lambda F} \ P(\lambda^{\text{out}})  \label{Eq_RevExtProcess}
\end{equation}
for some probability distribution $P(\lambda^{\text{out}})$. 
It follows from the fact that $\kappa^g_{X_1...X_n \lambda F}$ is a classical process that marginalization as in Eq.~\eqref{Eq_RevExtProcess} has to yield a classical process over nodes $X_1,...,X_n$ for arbitrary distributions $P(\lambda^{\text{out}})$, in particular for every point-distribution. 
Hence, for every value $\lambda'$ of $\lambda^{\text{out}}$, the induced function $g_{\lambda'} ( \_\_) :=g( \_\_ ,\lambda')$ has to define a deterministic process for $n+1$ nodes and furthermore, also once marginalizing over $F$ it still has to be a deterministic process for the $n$ nodes $X_1,\ldots , X_n$. Hence, Eq.~\eqref{Eq_RevExtProcess} can be read as establishing that the given $\kappa_{X_1...X_n}$ is a convex mixture of deterministic processes over the nodes $X_1, ..., X_n$, i.e. $\kappa_{X_1...X_n}$  lies in the deterministic polytope. 

Conversely, suppose $\kappa_{X_1...X_n}$ lies inside the deterministic polytope, that is, there exists a family of deterministic processes $\{\kappa^{f_i}_{X_1...X_n}\}_{i=1}^m$, defined by the functions $f_i : X_1^{\text{out}} \times ... \times X_n^{\text{out}} \rightarrow  X_1^{\text{in}} \times ... \times X_n^{\text{in}}$ such that 
$\kappa_{X_1...X_n} = \sum_{i=1}^m q_i \ \kappa^{f_i}_{X_1...X_n}$ for some probability distribution $\{q_i\}$. 
The proof will proceed by first observing that such a process can be seen to arise from one single deterministic process on $n+2$ nodes. Together with the fact that every deterministic process is reversibly extendible, proven in Ref.~\cite{Baumeler_2017_ReversibleTimeTravelWithFreedomOfChoice}, this establishes the claim. 
In order to see that indeed an appropriate deterministic process on $n+2$ nodes exists, let $\lambda^{\text{out}}$ and $F^{\text{in}}$ be variables with cardinality $m$ and define the function 
\begin{eqnarray}
	f & : & X^{\text{out}} \times \lambda^{\text{out}} \ \rightarrow \  X^{\text{in}} \times F^{\text{in}}  \\
	&& (x, \ i) \mapsto  (f_i(x) ,  \ i) \ , 
\end{eqnarray}
where $X^{\text{out}} = X_1^{\text{out}} \times ... \times X_n^{\text{out}}$ (similarly for $X^{\text{in}}$) and $x=(x_1,...,x_n)$.
Together with setting $P(\lambda^{\text{out}}=i) := q_i$, $f$ defines a deterministic classical process over the nodes $X_1,...,X_n$, $\lambda$ and $F$, which gives back $\kappa_{X_1...X_n}$ upon marginalization over $\lambda$ and $F$. That $f$ indeed defines a process follows from the fact that arbitrary variation of the distribution $P(\lambda^{\text{out}})$ corresponds to an arbitrary weighting $\{q_i\}$ in the originally given mixture, each case of which has to be a classical process.  
This concludes the proof.

\section*{ACKNOWLEDGEMENTS}

After completion of this work we became aware of a related result by Wataru Yokojima, Marco T{\'u}lio Quintino, Akihito Soeda and Mio Murao, which was obtained independently and has appeared in Ref.~\cite{YokojimaEtAl_2020_ConsequencesOfPreservingReversibilityInSupermaps} since the first preprint of this paper in Ref.~\cite{BarrettEtAl_2020_CyclicQCMs_FirstArxivVersion}. 
This work was supported by the EPSRC National Quantum Technology Hub in Networked Quantum Information Technologies, the Wiener-Anspach Foundation, and by the Perimeter Institute for Theoretical Physics. Research at Perimeter Institute is supported by the Government of Canada through the Department of Innovation, Science and Economic Development Canada and by the Province of Ontario through the Ministry of Research, Innovation and Science. 
This publication was made possible through the support of the ID\# 61466 grant from the John Templeton Foundation, as part of the “The Quantum Information Structure of Spacetime (QISS)” Project (qiss.fr). The opinions expressed in this publication are those of the authors and do not necessarily reflect the views of the John Templeton Foundation. 
This work was supported by the Program of Concerted Research Actions (ARC) of the Universit\'{e} libre de Bruxelles. O. O. is a Research Associate of the Fonds de la Recherche Scientifique (F.R.S.–FNRS).



\providecommand{\href}[2]{#2}\begingroup\raggedright\endgroup

\end{document}